\documentclass[12pt]{article}

\usepackage{mheck}

\newcommand{\rem}[1]{} 

\def\C{\mathbb{C}}
\def\Z{\mathbb{Z}}

\def\R{\mathbb{R}}

\def\P{\mathbb{P}}

\def\Hirz[#1]{\mathbbm{F}_{#1}}
\def\o[#1]{\overline{#1}}

\makeatletter
\newcommand\xleftrightarrow[2][]{%
  \ext@arrow 9999{\longleftrightarrowfill@}{#1}{#2}}
\newcommand\longleftrightarrowfill@{%
  \arrowfill@\leftarrow\relbar\rightarrow}
\makeatother

\institution{OXFORD}{\ Rudolf Peierls Centre for Theoretical Physics, Oxford University, Oxford, OX1 3NP, UK }
\institution{SCGP}{\ Simons Center for Geometry and Physics, SUNY, Stony Brook, NY, 11794-3636 USA}

\title{Mirror Symmetry for $G_2$-Manifolds:\\ Twisted Connected Sums and Dual Tops}
\authors{Andreas P. Braun \worksat{\OXFORD} \footnote{e-mail: {\tt andreas.braun@physics.ox.ac.uk}} and Michele Del Zotto \worksat{\SCGP} \footnote{e-mail: {\tt mdelzotto@scgp.stonybrook.edu}}}

\abstract{Recently, at least 50 million of novel examples of compact $G_2$ holonomy manifolds have been constructed as twisted connected sums of asymptotically cylindrical Calabi-Yau threefolds. The purpose of this paper is to study mirror symmetry for compactifications of Type II superstrings in this context. We focus on $G_2$ manifolds obtained from building blocks constructed from dual pairs of tops, which are the closest to toric CY hypersurfaces, and formulate the analogue of the Batyrev mirror map for this class of $G_2$ holonomy manifolds, thus obtaining several millions of novel dual superstring backgrounds. In particular, this leads us to conjecture a plethora of novel exact dualities among the corresponding 2d $\cn=1$ sigma models.}
\begin{document}

\maketitle

\tableofcontents

\section{Introduction}
Dualities along the landscape of superstring compactifications are one of the most important features of string theory. Among the known dualities, mirror symmetry for compactifications of Type II superstrings on Calabi-Yau (CY) manifolds is one of the most powerful. One of its manifestations, in the context of perturbative string theories at large volume, is the statement that two-dimensional sigma-models with different CY targets are related to exactly marginal deformations of the same two-dimensional SCFT \cite{Dixon:1987bg, Lerche:1989uy, Candelas:1989hd, Greene:1990ud}. The full quantum duality is expected to be even deeper than that, giving rise to an isomorphism for the whole quantum physics of the mirror compactifications \cite{Strominger:1995cz,Aspinwall:1995td,Morrison:1995yi,Strominger:1996it,Becker:1996ay}.

\begin{figure}
$$\xymatrix{\text{IIA}_J\,\,  \ar@{<->}[rr]^{\mu}\ar@{<..>}[dd]_{\nu}&&\,\,\text{IIA}_{J^\vee} \ar@{<..>}[dd]^{\nu}\\
\\
\text{IIB}_J\,\,  \ar@{<->}[rr]_{\mu}&&\,\,\text{IIB}_{J^\vee} }$$
\caption{Dualities for type II superstrings on a mirror pair of $G_2$ holonomy manifolds $(J,J^\vee)$.}\label{WTF}
\end{figure}
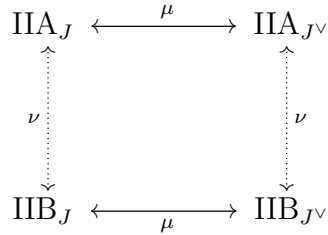

Similar dualities have been conjectured for manifolds with holonomy $G_2$ \cite{Shatashvili:1994zw,Papadopoulos:1995da,Acharya:1996fx, Acharya:1997rh}, giving rise to a network of generalized mirror symmetries --- see Figure \ref{WTF}. Let $J$ be a manifold with $G_2$ holonomy, and let $J^\vee$ denote its $G_2$-mirror. The vertical $G_2$-mirror map $\nu$ in Figure \ref{WTF} has been conjectured based on the fact that the compactifications of the Type IIA and Type IIB supergravities on $J$ agree \cite{Papadopoulos:1995da}. The horizontal $G_2$-mirror map $\mu$, which is going to be the focus of the present note, arises from four T-dualities via a generalization of the SYZ argument \cite{Acharya:1997rh}. Similar conjectures were originally formulated in the context of appropriate 2d extended $\cn=1$ SCFTs describing strings propagating on $G_2$-holonomy manifolds \cite{Shatashvili:1994zw} (see also \cite{Figueroa-OFarrill:1996tnk}).\footnote{ See also \cite{Howe:1991im,Howe:1991vs,Howe:1991ic,Howe:1994tv} for the corresponding sigma-models.} As in the CY case \cite{Odake:1988bh}, the origin of mirror symmetry from the CFT perspective is the presence of a non-trivial mirror automorphism of the (right moving) extended $\cn=1$ algebra \cite{Becker:1996ay,Roiban:2002iv,Gaberdiel:2004vx}. In particular, the case of the Joyce $T^7/(\mathbb{Z}_2)^3$ orbifolds has been analysed in detail in \cite{Gaberdiel:2004vx}, where each map in Figure \ref{WTF} was given an elegant interpretation in these terms. The moduli spaces of such 2d $\cn=1$ SCFTs are typically larger than the geometric moduli spaces usually considered by mathematicians, analogue to what happens for the K\"ahler moduli spaces of CYs \cite{Shatashvili:1994zw}. In the case of $G_2$ special holonomy, the 2d theories have a conformal manifold of dimension $b_2 + b_3 = b_2 + b_4$, where $b_n$ denotes the $n$-th Betti number of the manifold. In particular, $G_2$-mirror pairs must satisfy the Shatashvili-Vafa relation \cite{Shatashvili:1994zw}
\begin{equation}\label{SV}
b_2(J) + b_3(J) = b_2(J^\vee)+b_3(J^\vee).
\end{equation}
Notice that this agrees with what is expected from the reduction of the 9+1 dimensional Type II supergravities to 2+1 dimensions on a $G_2$ holonomy manifold preserving 4 supercharges. In particular, the (K\"ahler) metric on such moduli spaces should correspond to the Zamolodchikov metric \cite{Zamolodchikov:1986gt} on the conformal manifold for the 2d $\cn=1$ theory, analogously to what happens in the CY case \cite{Candelas:1989qn}.

Recently, lots of progress has been made in producing examples of compact $G_2$ holonomy manifolds. Indeed, at least 50 million can be easily generated by means of  twisted connected sums (TCS) of asymptotically cylindrical CY three-folds, following \cite{MR2024648,Corti:2012kd}. The physical implications of this fact are stunning.\footnote{ See e.g. \cite{Halverson:2014tya,Halverson:2015vta} for a discussion in the context of M-theory.} In particular, it is very natural to ask about the $G_2$-mirror map in this context.

In the CY case, the largest class of examples of Calabi-Yau manifolds for which a mirror is readily constructed (and in fact the largest class of examples of CY manifolds) is given by CY hypersurfaces and complete intersections in toric varieties \cite{Batyrev:1994hm,Batyrev:1994pg}. This construction rests on the polar duality between reflexive polytopes. A similar structure is in place for asymptotically cylindrical CY threefolds, whenever these are built from dual pairs of tops \cite{Braun:2016igl}. These give rise to $G_2$ manifolds in the TCS construction, and we claim that for this class of examples a structure analogous to that of the Batyrev mirror map is in place: the $G_2$-mirror pairs are canonically obtained by switching the roles of the dual tops used in the construction. We present an heuristic derivation of the above conjecture and a preliminary consistency check by verifying that the pairs of $G_2$ holonomy manifolds so obtained indeed satisfy the Shatashvili-Vafa relation. Our method allows to construct several millions of examples of such pairs, and we have discussed some explicitly to illustrate the power of the method.

In the case of CY 3-folds, mirror symmetry entails, in particular, the isomorphism of the lattices $H^{even}(X,\mathbb{Z})$ and $H^{odd}(X^\vee,\mathbb{Z})$ \cite{Aspinwall:1990ck,Gross:1997hn,Batyrev:2005jc}. It is natural to expect an analogous phenomenon in the context of $G_2$-mirrors. Indeed, as detailed in Section \ref{sect:mirrorgluing}, our construction directly yields the analogous relation
\begin{equation}\label{latticemir}
H^2(J,\mathbb{Z}) \oplus H^4(J,\mathbb{Z}) \simeq H^2(J^\vee,\mathbb{Z}) \oplus H^4(J^\vee,\mathbb{Z})
\end{equation}
for a $G_2$ mirror pair. Furthermore, we also expect the torsion in $H^2(J,\mathbb{Z}) \oplus H^3(J,\mathbb{Z})$ to be preserved. A thorough exploration of this, as well as its physical significance, is left for future work. 

A consequence of our conjecture is that we are providing examples of several millions of dual 2d $\cn=1$ sigma models. It would of course be extremely interesting to find an understanding of this duality from the 2d perspective, along the lines of e.g. \cite{Witten:1993yc,Hori:2000kt,Aganagic:2002mp}, and to study the interplay of this duality with topological $G_2$ strings \cite{deBoer:2005pt}. Another interesting angle is given by the geometric engineering perspective. M-theory compactifications on $J$ and $J^\vee$ lead to inequivalent 4d $\cn=1$ theories which become equivalent only upon circle reduction, mapping M theory to IIA. In the examples we consider, we have only abelian gauge groups and this is related to the fact vectors in 3d can be dualized to scalars. It would be very interesting to extend the $G_2$-mirror map to include  examples of $G_2$ manifolds in which we have more interesting gauge groups and matter contents \cite{Acharya:2000gb,Atiyah:2001qf,Witten:2001uq,Acharya:2001gy,Gukov:2002es,Acharya:2004qe}, perhaps along the lines of \cite{Aganagic:2001ug}. A further direction which we leave for future work is the relation among the duality discussed below and mirror symmetry for CYs --- see Remark 1 in \cite{Gukov:2002jv}.

This paper is organized as follows. In Section \ref{AllGore} we discuss an heuristic argument for the $G_2$-mirror symmetry for TCS $G_2$-manifolds based on $T$-duality, following \cite{Strominger:1996it,Acharya:1997rh}. In particular, this entails that the two asymptotically cylindrical CYs are swapped with their mirrors, and gives a rationale for the structure to be found. In Section \ref{sect:mirdualtops}, we review certain aspects of \cite{Braun:2016igl}, introduce the class of models which are going to be the focus of the present paper and formulate our conjecture. In Section \ref{EXAMPLS} we discuss some examples, to illustrate the power of the method. Technical details and proofs can be found in the Appendix.

\section{Mirror symmetry for TCS: heuristics}\label{AllGore}
To fix notation, let us begin with a quick informal review of the construction of TCS $G_2$-holonomy manifolds \cite{MR2024648,Corti:2012kd,MR3109862}. Consider a pair $X_+$ and $X_-$ of CY threefolds which are asymptotically cylindrical, meaning that they have one end which asymptotically has the form $\mathbb{R}^+ \times S^1 \times S_{\pm}$, where $S_{\pm}$ are smooth K3 surfaces.\footnote{For a precise definition see \cite{MR3109862}, Definition 2.4.} In particular, the metric, K\"ahler form, and holomorphic top form on the asymptotic CY cylinders converge to
\begin{equation}
ds_{\pm}^2 = dt^2 + d\theta^2 + ds^2_{S_{\pm}}, \qquad \omega_{\pm} =  dt \wedge d\theta + \omega_{S_{\pm}}, \qquad \Omega^{3,0}_{\pm} = (d\theta - i 
dt)\wedge \Omega^{2,0}_{S_{\pm}}, 
\end{equation}
in obvious notation. Now consider the products $S^1 \times X_+$ and $S^1 \times X_-$. Each side can be equipped with a $G_2$-structure
\begin{equation}\label{assG2}
\varphi_{\pm} \equiv d \xi \wedge \omega_{\pm} + \text{Re}(\Omega^{3,0}_{\pm}) \qquad \star \varphi_{\pm} \equiv \tfrac{1}{2} \omega_{\pm}^2 - d\xi \wedge \text{Im} (\Omega_{\pm}^{3,0}),
\end{equation} 
where we have denoted by $\xi$ the coordinate of the extra $S^1$. Consider the asymptotically cylindrical regions, fix an $\ell >0$ large enough, and let $t \in (\ell, \ell+1) \subset \mathbb{R}^+$. Consider the diffeomorphism: $\Xi_\ell\colon S^1 \times X_+ \to S^1 \times X_-$, which in local coordinates is given by
\begin{equation}\label{THADIFFEOH}
\Xi\colon (\xi, t , \theta, Z) \mapsto (\theta, \ell+1 - t , \xi, g(Z)),
\end{equation}
where $g: S_+ \to S_-$ is a hyperk\"ahler rotation, i.e. a diffeomorphism of $K3$ surfaces which induces
\begin{equation}\label{HKAROTAH}
\begin{aligned}
&g^*ds^2_{S_-} = ds^2_{S_+}, & g^*\text{Im}(\Omega^{2,0}_{S_-})= - \text{Im}(\Omega^{2,0}_{S_+}), \\
&g^*\text{Re}(\Omega^{2,0}_{S_-})=\omega_{S_+}, &g^*\omega_{S_-}=\text{Re}(\Omega^{2,0}_{S_+}).
\end{aligned}
\end{equation}
This is called a matching in \cite{Corti:2012kd}. Notice that from the definition follows that
\begin{equation}\label{THAGLUAH}
\Xi^*\varphi_- \equiv \varphi_+.
\end{equation}
Truncating both manifolds $S^1 \times X_{\pm}$ at $t = \ell+1$ one obtains compact manifolds with boundaries $S^1 \times S^1 \times S_{\pm}$ which can be glued via the diffeomorphism $\Xi_\ell$. By Theorem 3.12 of \cite{Corti:2012kd}, for sufficiently large $\ell$, the manifold $J$ so obtained is a $G_2$-holonomy manifold.

A beautiful geometrical approach to $G_2$-mirror symmetry is given by generalizing the SYZ argument to $G_2$-holonomy manifolds \cite{Acharya:1997rh}. The $G_2$-holonomy manifolds have two natural classes of calibrated submanifolds, associative submanifolds, which are calibrated by the 3-form $\varphi$, and coassociative submanifolds, which are calibrated by $\star\varphi$ \cite{Harvey:1982xk}. Deformations of associative submanifolds are obstructed, while deformations of coassociative ones are not: a coassociative submanifold $N$ has a smooth moduli space of dimension $b_2^+(N)$, the number of self-dual harmonic 2-forms \cite{Mclean96}. Let $(J,J^\vee)$ denote a putative $G_2$-mirror pair. In a compactification of IIA on $J$, a D$0$-brane has a moduli space which equals $J$, which must correspond to the moduli space of a wrapped D$p$-brane on $J^\vee$. As we want a BPS configuration, the only option left is wrapping a coassociative $N\subset J^\vee$ with a D4-brane. The $U(1)$ vector field on the brane gives rise to $b_1(N)$ additional moduli, whence the physical moduli space has dimension $b_1(N)+b_2^+(N)$. For this to coincide with the D$0$ brane the moduli spaces must agree, whence $b_1(N)+b_2^+(N) = 7$. It is hence natural to conjecture that $N\simeq T^4$ \cite{Acharya:1997rh}. In what follows we are going to argue that this is indeed the case for the TCS $G_2$ manifolds. Four T-dualities along the cycles of such a $T^4$ map the D4-brane on $J^\vee$ back to the D0-brane on $J$, so that repeating the argument vice-versa this entails that $J$ has an analogous $T^4$ fibration. This realizes the $G_2$-mirror map $\mu$ in Figure \ref{WTF} as four T-dualities along such a coassociative $T^4$.

Let us proceed with our heuristic argument about $G_2$-mirror symmetry for such $J$. Consider the mirrors $X^{\vee}_{\pm}$ of the asymptotically cylindrical Calabi-Yau manifolds $X_{\pm}$ and let $L_\pm$ be the corresponding SYZ special lagrangian $T^3$ \cite{Strominger:1996it}. In the asymptotically cylindrical region of the manifold  $X_{\pm} \sim \mathbb{R}^+ \times S^1 \times S_{\pm}$, the SYZ special lagrangians must asymptote to $L_\pm \sim S^1 \times \Lambda_\pm$, where $\Lambda_{\pm}$ are special lagrangian $T^2$ within the asymptotic K3s with respect to the K3 complex structure induced by the ambient CY. In particular, they do not extend along the $\mathbb{R}^+$ direction. Let us choose the holomorphic top form on $X_\pm$ such that
\begin{equation}\label{mademeup}
- i \, \text{vol}_{L_\pm} = \Omega_{\pm}^{3,0}|_{L_\pm}.
\end{equation}
Notice that from Eqn.\eqref{assG2} a special lagrangian $L$ satisfying \eqref{mademeup} always gives rise to a coassociative cycle $N_L \equiv S^1 \times L \subset S^1 \times X$. In particular,
\begin{equation}
\text{Im} (\Omega_{\pm}^{3,0}) = d \theta \wedge \text{Im}(\Omega^{2,0}_{S_{\pm}}) - dt \wedge \text{Re}(\Omega^{2,0}_{S_{\pm}})
\end{equation}
therefore, for our special lagrangians $L_\pm$ we have
\begin{equation}
\text{Im} (\Omega_{\pm}^{3,0})|_{L_\pm} = d \theta \wedge \left(\text{Im}(\Omega^{2,0}_{S_{\pm}})|_{\Lambda_\pm}\right),
\end{equation}
and by tacking the coassociatives $ N_\pm \equiv S^1 \times L_\pm \subset S^1 \times X_\pm$ we get
\begin{equation}
\star \varphi_\pm|_{N_\pm} = - d \xi \wedge d\theta \wedge \left(\text{Im}(\Omega^{2,0}_{S_{\pm}})\right)|_{\Lambda_{\pm}}.
\end{equation}
In particular
\begin{equation}
\Xi^*(\star\,\varphi_-|_{N_-}) = \star\,\varphi_+|_{N_+}
\end{equation}
which follows by swapping the two $S^1$s and changing sign of $\text{Im}(\Omega^{2,0})$ as dictated by the hyperk\"ahler rotation in Eqn.\eqref{HKAROTAH}. Therefore, the twisted connected glueing diffeomorphism $\Xi$ is also glueing $N_\pm$ to a coassociative submanifold $M\subset J$ which has the topology of a $T^4$ that may become singular along loci in $J$. Performing three T-dualities along the $L_\pm$ SYZ fibres is mapping $X_{\pm}$ to their mirrors $X_{\pm}^\vee$ by construction. However, as the cycles of the asymptotic cylinders are swapped with the extra $S^1$'s along the glueing, they must have the same size and we have to necessarily perform four T-dualities along the $T^4$ coassociative $M$. The resulting manifold is the $G_2$-mirror $J^\vee$ of $J$.

Notice that by construction $X^\vee_\pm$ are asymptotically cylindrical as well. We claim that $J^\vee$ is itself a twisted connected sum obtained out of the CY mirrors $X^\vee_\pm$ of $X_\pm$. In order to show this, the only thing left to do is to discuss how the original hyperk\"ahler rotation $g$ transforms under $G_2$-mirror symmetry. Notice that in the asymptotically cylindrical region where the twisted connected sum occurs, we see that two of the four T-dualities occur along the $\Lambda_\pm$ special lagrangians within the smooth asymptotic K3 surfaces $S_\pm$, thus inducing mirror symmetries on the asymptotic K3 fibres in the glueing region. The asymptotic cylinders of the mirror $X_\pm^{\vee}$ have the form $\mathbb{R^+} \times (S^1)^\vee \times S_\pm^\circ$, where $S\to S^\circ$ is the K3 mirror map as defined e.g. in Section 3.4 of \cite{Aspinwall:1996mn} (see Appendix \ref{sect:mirrorfork3} for a review). In fact, compatibility with the K3 mirror symmetry suggests to extend the action of the hyperk\"ahler rotation in Eqn.\eqref{HKAROTAH} to the $B$ field on K3 as follows
\begin{equation}
g^*B_- = - B_+,
\end{equation}
so  that a canonical $g^\vee$ can be obtained by the composition
\begin{equation}\label{dualg}
g^\vee \equiv S^\circ_+ \xrightarrow{\quad mir \quad } S_+ \xrightarrow{\quad  g \quad } S_- \xrightarrow{\quad  mir \quad } S^\circ_-\,.
\end{equation}
Let us proceed by checking that the $g^\vee$ so defined indeed gives a hyperk\"ahler rotation for the pair $X^\vee_\pm$. The mirror map $S_\pm \leftrightarrow S^\circ_\pm$ gives, in particular
\begin{equation}
(\omega_\pm,B_\pm,\text{Re}( \Omega_\pm),\text{Im}( \Omega_\pm)) \longleftrightarrow (\text{Re}( \Omega^\circ_\pm),\text{Im}( \Omega^\circ_\pm),\omega^\circ_\pm,B^\circ_\pm)
\end{equation}
therefore, the chain of maps in Eqn.\eqref{dualg} reads:
\begin{equation}
\begin{aligned}
&\omega^\circ_+ \mapsto \text{Re}(\Omega_+) \mapsto \omega_- \mapsto \text{Re}(\Omega^\circ_-)\\
&\text{Re}(\Omega_+^\circ) \mapsto \omega_+\mapsto\text{Re}(\Omega_-) \mapsto \omega^\circ_-\\
&\text{Im}(\Omega_+^\circ) \mapsto B_+ \mapsto - B_- \mapsto - \text{Im}(\Omega^\circ_-),
\end{aligned}
\end{equation}
and indeed $g^\vee:S^\circ_+ \to S^\circ_-$ is a hyperk\"ahler rotation, as desired. This concludes our heuristic argument showing that $J^\vee$ is indeed a $G_2$-holonomy manifold obtained as a twisted connected sum of the pair $(X^\vee_+,X^\vee_-)$, which are the CY mirrors of $(X_+,X_-)$. Of course, there are lots of subtleties we are not addressing here (which are in part related with the subtleties in the original SYZ argument \cite{Morrison:2010vf,Gross:2012rw} and also go beyond), but this argument is meant to be no more than a motivation to look for TCS $G_2$-mirror pairs $(J,J^\vee)$ with such a structure. Remarkably, such a structure naturally emerges for asymptotically cylindrical Calabi-Yau threefolds constructed from dual tops \cite{Braun:2016igl}.

\section{Mirror symmetry for $G_2$-manifolds from dual tops}\label{sect:mirdualtops}

\subsection{Building Blocks}\label{sect:defsbuildbocks}
Building blocks are threefolds which give a remarkably elegant way of producing the asymptotically cylindrical CYs needed in the TCS construction of $G_2$-manifolds \cite{MR2024648,Corti:2012kd,MR3109862}. A building block $Z$ is fibration $\pi: Z \to \P^1$ (whose generic fibre $\pi^{-1}(p) \equiv S_p$ is a non-singular K3 surface) with the further properties that \cite{MR2024648,Corti:2012kd,MR3109862}:
\textit{i.)} the anticanonical class of $Z$ is primitive\footnote{ This means that 
there is no line bundle $L$ such that $L^{\otimes n} =  [K_Z]$ 
for any $n > 1$.} and equal to the class of the generic fibre, $S$: $[-K_Z] = [S]$; \textit{ii.)} we may pick a smooth and irreducible fibre $S_0$, such that there is no monodromy upon orbiting around $S_0$, i.e. the fibration is trivial in the vicinity of $S_0$. There is a natural restriction map 
\begin{equation}\label{eq:defN}
\rho: H^2(Z,\mathbb{Z}) \rightarrow H^2(S_0,\mathbb{Z}) \cong \Gamma^{3,19}  = (-E_8^{\oplus 2}) \oplus U^{\oplus 3}\, ;
\end{equation}
\textit{iii.)} Denoting the image of $\rho$ by $N$, we demand that the quotient $\Gamma^{3,19} /N$ is torsion free, i.e. the embedding $N \hookrightarrow \Gamma^{3,19} $ is primitive; and \textit{iv.)} $H^3(Z,\mathbb{Z})$ has no torsion. Under these assumptions, it follows that $Z$ is simply connected and the Hodge numbers $H^{1,0}(Z)$ and $H^{2,0}(Z)$ vanish. As $Z$ is a $K3$ fibration over
$\P^1$, the normal bundle of the fibre, and in particular of $S_0$, is trivial. The lattice $N$ naturally embeds into the Picard lattice of 
$S_0$ and we can think of the fibres as being elements of a family of lattice polarized K3 surfaces with polarizing lattice $N$. By excising a fibre, we may form the open space 
\begin{equation}
 X \equiv Z \setminus S_0.
\end{equation}
The manifold $X$ so obtained is an asymptotically cylindrical CY threefold \cite{Corti:2012kd}. The data defining the pair of building blocks $Z_{\pm}$ corresponding to $X_{\pm}$ is enough to reconstruct the homological properties of the corresponding TCS $G_2$-manifold. We summarize some relevant formulas from \cite{Corti:2012kd} in appendix \ref{FORMULAHOM}.

Let us remark that one may think of the $J$ so obtained as a (non-holomorphic) K3 fibration over a three-dimensional base. Such a base is furthermore a fibration of a torus over an interval for which one of the two circles of the torus collapses at each end: using Hopf coordinates on $S^3$, one can see that this space is indeed topologically a 3-sphere.\footnote{ See, e.g., Figure 1 of \cite{Earp:2013jea}.} This has interesting consequences for the physics of these models which we will explore elsewhere.

\subsection{Building Blocks from Projecting Tops} \label{sect:tops_as_build}

A pair of lattice polytopes $(\Delta,\Delta^\circ)$ satisfying
\begin{equation}
\langle \Delta, \Delta^\circ \rangle \geq -1 \, 
\end{equation}
under the canonical pairing on $\R^n$ are called reflexive and define a Calabi-Yau manifold $X_{(\Delta,\Delta^\circ)}$ embedded as a hypersurfaces in a toric variety \cite{Batyrev:1994hm}. In this construction, the polytope $\Delta$ is the Newton polytope giving rise to all of the monomials of the defining equation and the polytope $\Delta^\circ$, after an appropriate triangulation, defines the toric ambient space. Crucially, the normal fan of the polytope $\Delta$ is equivalent to the fan over the faces of $\Delta^\circ$, which allows for a derivation of simple combinatorial formulas for the Hodge numbers of $X_{(\Delta,\Delta^\circ)}$ using the techniques of \cite{MR873655}. 

In a similar fashion, the building blocks used in the construction of $G_2$ manifolds as twisted connected sums can be obtained from a pair of four-dimensional projecting tops $\Diamond, \Diamond^\circ$ \cite{Braun:2016igl}. A top $\Diamond^\circ$ is defined as a bounded lattice polytope (w.r.t. a lattice ${\bf N}$) defined by relations
\be
\begin{aligned}
\langle m_i, \Diamond^\circ &\rangle \geq -1 \\
\langle m_0, \Diamond^\circ &\rangle \geq 0 
\end{aligned}
\ee
for a set of (primitive) lattice points $m_i$ and $m_0$, all sitting in the dual lattice ${\bf M}$. The last relation defines a hyperplane $F$ and $\Diamond^\circ \cap F$ must be a reflexive polytope $\Delta^\circ_F$. Tops appear naturally as halves of reflexive polytopes defining Calabi-Yau hypersurfaces which are fibred by a Calabi-Yau hypersurface of one dimension lower, which is in turn defined by the reflexive pair $(\Delta_F,\Delta^\circ_F)$ \cite{Klemm:1995tj,Candelas:1996su,Avram:1996pj,Kreuzer:2000qv}. Let us specialize to our case of interest, in which ${\bf N}$ and ${\bf M}$ are four-dimensional. We may always exploit $SL(4,\mathbb{Z})$ to fix $m_0 = (0,0,0,1)$, and, following \cite{Candelas:2012uu}, a top with this choice of $m_0$ is called projecting if the projection $\pi_4$ forgetting the fourth coordinate maps $\pi_4(\Diamond^\circ)\supseteq \Delta^\circ_F$. 

For any projecting top $\Diamond^\circ$ with $\Diamond^\circ\cap F =\Delta^\circ_F$, $F = m_0^\perp$, there is a dual top $\Diamond$ satisfying:
\begin{equation}\label{eq:topsduality}
 \begin{aligned}
 & \langle \Diamond, \Diamond^\circ \rangle \geq -1 & \\
  \langle  \Diamond,\nu_0 \rangle \geq 0 \hspace{.5cm} & &  \langle m_0, \Diamond^\circ\rangle \geq 0
\end{aligned}
\end{equation}
with $\nu_0 = (0,0,0,-1)$. Here, our notation $^\circ$ is meant to indicate `dual' in the sense of the above relation rather than `polar dual'.
As a convex lattice polytope, $\Diamond$ defines a toric variety $\P_{\Sigma_n(\Diamond)}$ via a normal fan $\Sigma_n(\Diamond)$, as well as a line bundle $\mathcal{O}(D_\Diamond)$ on $\P_{\Sigma_n(\Diamond)}$. Contrary to the case of reflexive pairs, the face fan $\Sigma_f(\Diamond\cup\nu_0)$ of $\Diamond^\circ \cup \nu_0$ is in general not equal to, but a refinement of $\Sigma_n(\Diamond)$ \cite{Braun:2016igl}. 

A generic section of $\mathcal{O}(D_\Diamond)$ defines a hypersurface $Z_s$ and $\P_{\Sigma_f(\Diamond\cup\nu_0)}$ may have singularities which meet $Z_s$. Similar to the case of reflexive polytopes, one can further refine the fan $\Sigma_f(\Diamond\cup\nu_0)$ according to a (fine, star, projective) triangulation of $\Diamond^\circ$ to find a maximally crepant desingularisation. In our case of interest, where $Z$ is a threefold and $\P_\Sigma$ a fourfold, such a triangulation will only leave point-like singularities in $\P_\Sigma$ which do not meet a generic hypersurface.\footnote{ The reason for this is that any fine triangulation of a face of dimension less than three leads to simplices of lattice volume unity.} The smooth hypersurface $Z_{(\Diamond,\Diamond^\circ)}$ after resolution is then given by
\begin{equation}\label{eq:zdefeq}
Z_{(\Diamond,\Diamond^\circ)}: \,\,\, 0 = \sum_{m \in \Diamond} z_0^{\langle m, \nu_0 \rangle}\prod_{\nu_i} z_i^{\langle m, \nu_i\rangle +1} 
\end{equation}

For a projecting top, $\Delta^\circ_F = \Diamond^\circ \cap F$ and $\Delta_F = \Diamond \cap F$ are a reflexive pair \cite{Avram:1996pj}. The hypersurface
given by the vanishing locus of a section of $\mathcal{O}(D_\Diamond)$, which we denote by $Z$, is fibred by a $K3$ surface which is defined by the reflexive pair $(\Delta_F, \Delta_F^\circ)$. 

There is an intuitive way to think about the building blocks $Z_{(\Diamond,\Diamond^\circ)}$ as resulting from a degeneration of a $K3$ fibred Calabi-Yau threefold. Let us assume that we are given two tops $\Diamond^\circ_a$ and $\Diamond^\circ_b$ which share the same $\Delta_F^\circ$. These may be combined to form a reflexive polytope $\Delta^\circ$ \cite{Avram:1996pj,Candelas:2012uu}, which in turn defines a family $X_{(\Delta,\Delta^\circ)}$ of $K3$ fibred CY threefolds. As detailed in Appendix \ref{sect:degcy3tobb}, such CY threefolds have a degeneration limit in which they split into $Z_{(\Diamond_a,\Diamond^\circ_a)}$ and $Z_{(\Diamond_b,\Diamond^\circ_b)}$, with the two components intersecting along a $K3$ surface $X_{(\Delta_F,\Delta^\circ_F)}$. In this limit, the base $\P^1$ of $X_{(\Delta,\Delta^\circ)}$ becomes very long with the $K3$ fibre essentially constant (and equal to $S_0$) in the cylindrical region. If we cut $X_{(\Delta,\Delta^\circ)}$ along the $S^1$ of the cylinder in the `bulk' region, we find $Z_{(\Diamond_a,\Diamond^\circ_a)}\setminus S_{0}$ and $Z_{(\Diamond_b,\Diamond^\circ_b)}\setminus S_{0}$. We can hence think of $Z_{(\Diamond,\Diamond^\circ)}\setminus S_0$ as half a CY threefold. This degeneration limit generalizes the degeneration of an elliptic $K3$ surface into two rational elliptic surfaces ($dP_9$'s), which are a lower-dimensional analogue to the threefolds $Z_{(\Diamond,\Diamond^\circ)}$ considered here. 

Using the above construction, one can derive combinatorial formulas for the Hodge numbers of $Z_{(\Diamond,\Diamond^\circ)}$, as well as the (ranks of the) lattices 
\begin{equation}
\begin{aligned}
 N\left(Z_{(\Diamond,\Diamond^\circ)}\right) &= im(\rho)  \\
K\left(Z_{(\Diamond,\Diamond^\circ)}\right)  &= ker(\rho)/[S_0] \, ,
\end{aligned}
\end{equation}
which are given in appendix \ref{HpqNKs}.

\subsection{Mirror Building Blocks}\label{sect:mirror}

For a pair of reflexive polytopes, it is well-known that exchanging the roles played by $\Delta$ and $\Delta^\circ$ produces the mirror Calabi-Yau $X_{(\Delta^\circ,\Delta)} = (X_{(\Delta,\Delta^\circ)})^\vee$. Similarly, it is a natural operation to swap the dual pair of tops, i.e. swap the building blocks
\begin{equation}
Z = Z_{(\Diamond,\Diamond^\circ)} \hspace{.5cm} \leftrightarrow \hspace{.5cm} Z^\vee = Z_{(\Diamond^\circ,\Diamond)} \, .
\end{equation}
This reversing of the roles of the two tops imitates Batyrev's construction \cite{Batyrev:1994hm} of mirror pairs of Calabi-Yau threefolds. Correspondingly, we will call $Z,Z^\vee$ a mirror pair of building blocks. 

As suggested by the heuristics in Section \ref{AllGore}, notice that we may choose the constant fibres $S_0$ of the cylindrical region of $X = Z\setminus S_0$ to be mirror\footnote{ While it is clear that this can be done in the lattice polarized families, it is a subtle question if the corresponding points in moduli space are realized in the \emph{algebraic} families. We ignore this question in this work.} of the fibres $S_0^\vee$ of the cylindrical region of $X^\vee=Z^\vee \setminus S_0^\vee$, but that all other fibres will not be mirror (though being part of algebraic mirror families). Mirror symmetry swaps the K\"ahler form, which stays constant over the base, with the real part of $\Omega$, which varies over the base. Replacing a top with its mirror hence does \emph{not} correspond to fibre-wise mirror symmetry. This is very similar to the state of affairs for the large class of toric Calabi-Yau threefolds which are $K3$ fibred. 

The above discussion also gives us another insight into the nature of mirror symmetry of building blocks as derived from mirror symmetry for Calabi-Yau threefolds. Consider again a Calabi-Yau threefold $X_{(\Delta,\Delta^\circ)}$ for which $\Delta^\circ$ is formed of two projecting tops $\Diamond^\circ_a$ and $\Diamond^\circ_b$ which share the same $\Delta^\circ_F$. As discussed in detail in Appendix \ref{sect:degcy3tobb}, $X_{(\Delta,\Delta^\circ)}$ has a limit in which in which it degenerates into $Z_{(\Diamond_a,\Diamond^\circ_a)} \vee Z_{(\Diamond_b,\Diamond^\circ_b)}$. Equivalently, $X_{(\Delta,\Delta^\circ)}$ can be thought of as being glued from $Z_{(\Diamond_a,\Diamond^\circ_a)}\setminus S_{0}$ and $Z_{(\Diamond_b,\Diamond^\circ_b)} \setminus S_{0}$. In the degeneration limit, the $\P^1$ base of $X_{(\Delta,\Delta^\circ)}$ becomes stretched and all of the singular $K3$ fibres are localized close to the two poles. In the bulk region of the $\P^1$ base, which now looks like a cylinder, the fibre becomes constant and equal to $S_0$. Similarly, the mirror $X_{(\Delta^\circ,\Delta)}$ has a limit in which it degenerates into the mirror building blocks $Z_{(\Diamond^\circ_a,\Diamond_a)} \vee Z_{(\Diamond^\circ_b,\Diamond_b)}$, with the fibre in the bulk region of the $\P^1$ becoming the mirror $K3$ surface $S_0^\vee$. As $X_{(\Delta,\Delta^\circ)}$ and $X_{(\Delta^\circ,\Delta)}$ are related by performing three T-dualities along the SYZ fibres, which become a product of the cylinder $S^1$ with the SYZ fibre of $S_0$ in the bulk region of the base $\P^1$, it follows that $Z_{(\Diamond,\Diamond^\circ)}\setminus S_0$ and $Z_{(\Diamond^\circ,\Diamond)}\setminus S_0$ are mirror Calabi-Yau manifolds in the sense of SYZ. 

By carefully examining the combinatorial formulae of \cite{Braun:2016igl} one can show that (these relations are proved in Appendix \ref{sect:mirrorbuildblocksdetails}): 
\begin{enumerate}
 \item[a)] The lattices $N_{(\Diamond,\Diamond^\circ)}$ and $N_{(\Diamond^\circ,\Diamond)}$ admit a primitive embedding
 \begin{equation}\label{eq:dualrel1}
N_{(\Diamond,\Diamond^\circ)}\oplus N_{(\Diamond^\circ,\Diamond)} \oplus U \hookrightarrow \Gamma^{3,19}  \, ,
\end{equation}
where $\Gamma^{3,19} $ is the lattice $H^{2}(S,\mathbb{Z})$ of integral cycles of a $K3$ surface, i.e. the unique even unimodular lattice of signature $(3,19)$.\footnote{ Notice that this implies that a pair of $K3$ surfaces with lattice polarizations $N_{(\Diamond,\Diamond^\circ)}$ and $N_{(\Diamond^\circ,\Diamond)}$ form an algebraic mirror pair \cite{Aspinwall:1994rg,MR1420220}, generalizing the observation of \cite{Rohsiepe:2004st}.\label{THAFOOTNOTAH}} 
\item[b)] For a mirror pair of building blocks, the rank of the lattice $K$ and the Hodge number $h^{2,1}$ are swapped
\begin{equation}\label{eq:dualrel2}
\begin{aligned}
|K(Z_{(\Diamond,\Diamond^\circ)})| = h^{2,1}(Z_{(\Diamond^\circ,\Diamond)}) \\
|K(Z_{(\Diamond^\circ,\Diamond)})| = h^{2,1}(Z_{(\Diamond,\Diamond^\circ)}) 
\end{aligned} 
\end{equation}
\end{enumerate}

\subsection{Mirror $G_2$ Manifolds} \label{sect:mirrorgluing}
Let us now consider a $G_2$ manifold $J$ which is constructed as a twisted connected sum of two building blocks, which are in turn each obtained from a dual pair of tops, $Z_\pm = (\Diamond_\pm,\Diamond_\pm^\circ)$. Using \eqref{eq:dualrel2}, a glance at \eqref{eq:b2pb3fororthgluing} reveals that we can find many $G_2$ manifolds with the same $b_2 + b_3$ if we simply replace one building block (or both) by its mirror, while using arbitrary orthogonal gluing throughout. While this is certainly encouraging, it is not really what we want: our heuristic arguments of Section \ref{AllGore} imply that we are supposed to swap both $Z_\pm \setminus S_{0 \pm}$ with their mirrors. Also, we are looking for an operation of order two, corresponding to the automorphism in the 2d extended $\cn=1$ SCA of Shatashvili-Vafa \cite{Becker:1996ay,Roiban:2002iv,Gaberdiel:2004vx}. Our heuristic arguments further
imply that the $K3$ surfaces $S_{0 \pm}$ in the asymptotic cylinders should be replaced by their mirrors $S_{0 \pm}^\vee$. This fits nicely with relation a) above, which states that the fibres of the mirror building blocks $Z_{(\Diamond,\Diamond^\circ)}$ and $Z_{(\Diamond^\circ,\Diamond)}$ are from algebraic mirror families of $K3$ surfaces. The only ingredient missing is how to find a matching \eqref{HKAROTAH} between $S_{0 \pm}^\vee$ given one for $S_{0 \pm}$.

Recall that mirror symmetry for $K3$ surfaces includes a choice of $B$-field and takes place in the unique even self-dual lattice $\Gamma^{4,20}$. Here $\Gamma^{4,20}$ is decomposed as (see discussion in Appendix \ref{sect:mirrorfork3})
\begin{equation}
\Gamma^{4,20} = U_N \oplus U_T \oplus \Gamma^{2,18} \qquad\text{with}\qquad \Gamma^{2,18} \supseteq N \oplus \tilde{T},
\end{equation}
where $N$ is the polarizing lattice of the family and $\tilde{T}$ is its orthogonal complement in $\Gamma^{2,18}$. Under mirror symmetry
\begin{equation}
  N \,\leftrightarrow \,\tilde{T} \quad\text{and}\quad U_N \,\leftrightarrow U_T 
\end{equation}
are swapped. Let us now see the interplay of mirror symmetry for the $K3$ fibres $S_{0,\pm}$ with the gluing. A gluing is specified by primitive embeddings
\begin{equation}
N_\pm \hookrightarrow \Gamma^{3,19} 
\end{equation}
and a matching of the K\"ahler forms $\omega_\pm$ and the holomorphic two forms $\Omega_\pm$ in Eqn.\eqref{HKAROTAH} (see also Eqn.\eqref{eq:hkrot2}). In the light of mirror symmetry for K3 surfaces, we are interested in lifting this construction to $\Gamma^{4,20}$ and to include a $B$-field in the lattice $N$ of every K3 fibre constant over the base of a building block. From the perspective of mirror symmetry of K3 it is also natural to extend the definition of the hyperk\"ahler rotation in such a way that $g^*B_- = -B_+$ as discussed in Section \ref{AllGore}. As the lattices $N_\pm$ are only embedded into $\Gamma^{3,19}$, so that they stay orthogonal to $U_N$, and furthermore mirror symmetry swaps $U_N \leftrightarrow U_T$, it seems natural to consider embeddings for which $N_\pm$ also stay orthogonal to $U_T$. Let us hence consider a $G_2$ manifold $J$ constructed from two building blocks $Z_\pm$ and another $G_2$ manifold $J^\vee$ constructed from the mirrors building blocks $Z_\pm^\vee$. Here, we use an embedding where $N_\pm$ stay orthogonal to $U_T \oplus U_N$ together with the mirror matching $g^\vee$:
\begin{equation}\label{eq:mirrorglue}
\omega_\pm^\circ = Re(\Omega_\mp^\circ),  \qquad  Im(\Omega_+^\circ) =  - Im(\Omega_-^\circ), \quad\text{and}\quad B_+^\circ = - B_-^\circ,
\end{equation}
obtained from the matching data of $J$
\begin{equation}\label{eq:matchingdata}
\omega_\pm = Re(\Omega_\mp), \qquad Im(\Omega_+) = - Im(\Omega_-), \quad\text{and}\quad B_+ = - B_-.
\end{equation}
It follows from relation a) in Section \ref{sect:mirror} that
\begin{equation}
\tilde{T}_{(\Diamond,\Diamond^\circ)} =  N_{(\Diamond^\circ,\Diamond)} \, ,
\end{equation}
so that for such embeddings the only non-trivial contributions to $H^2(J,\Z)\oplus H^4(J,\Z)$ from Eqn.\eqref{eq:bettiG2} satisfy
\begin{equation}
\begin{aligned}
  & N_+ \cap N_- \oplus (T_+ \cap T_-) \oplus  \Gamma^{3,19} /(N_- + T_+) \oplus \Gamma^{3,19} /(N_+ + T_-) \\
= \,& N_+ \cap N_- \oplus (\tilde{T}_+ \cap \tilde{T}_-) \oplus U \oplus  \Gamma^{2,18} /(N_- + \tilde{T}_+) \oplus \Gamma^{2,18} /(N_+ + \tilde{T}_-) \\
= \,& \tilde{T}_+^\circ \cap \tilde{T}_-^\circ \oplus (N_+^\circ \cap N_-^\circ) \oplus U  \oplus  \Gamma^{2,18} /(\tilde{T}_-^\circ + N_+^\circ) \oplus \Gamma^{2,18} /(\tilde{T}_+^\circ + N_-^\circ) \, .
\end{aligned}
\end{equation}
As replacing both building blocks by their mirrors furthermore exchanges $h^{2,1}$ with $|K|$ by \eqref{eq:dualrel2}, it now follows from Eqn.\eqref{eq:bettiG2} that 
\begin{equation}
\begin{aligned}
b_2(J) + b_4(J) & = b_2(J^\vee) + b_4(J^\vee)  \\                                          
Tors(H^4(J,\Z)) & = Tors(H^4(J^\vee,\Z)) \, .
\end{aligned}
\end{equation}
Hence both the torsion subgroups and the Betti numbers agree, so that we can conclude
\begin{equation}\label{eq:equalityofh2h4}
H^2(J,\Z) \oplus H^4(J,\Z) \cong H^2(J^\vee,\Z) \oplus H^4(J^\vee,\Z) \, .
\end{equation}
Of course, this also implies that $b_2(J) + b_3(J) = b_2(J^\vee) + b_3(J^\vee)$ by Poincar\'e duality. 

\section{Examples}\label{EXAMPLS}

\subsection{Building Blocks fibred by a Quartic $K3$ surface}\label{sect:example_Quartic}

As the simplest algebraic realization of a $K3$ is given by a quartic hypersurface in $\P^3$, the simplest building block can be found as a hypersurface in $\P^3 \times \P^1$ of bidegree $(4,1)$. In the language of tops, this means we consider a pair of dual tops with vertices
\begin{equation}
\Diamond^\circ = \left(\begin{array}{rrrrr}
-1 & 0 & 0 & 0 & 1 \\
-1 & 0 & 0 & 1 & 0 \\
-1 & 0 & 1 & 0 & 0 \\
0 & 1 & 0 & 0 & 0
\end{array}\right) \, , \hspace{1cm} \Diamond =\left(\begin{array}{rrrrrrrr}
-1 & -1 & 3 & 3 & -1 & -1 & -1 & -1 \\
-1 & -1 & -1 & -1 & 3 & 3 & -1 & -1 \\
-1 & -1 & -1 & -1 & -1 & -1 & 3 & 3 \\
-1 & 0 & 0 & -1 & 0 & -1 & 0 & -1
\end{array}\right) 
\end{equation}
Adding the extra ray $\nu_0 = (0,0,0,-1)$ an applying \eqref{eq:zdefeq} the reproduces a hypersurface of bidegree $(4,1)$ in $\P^3 \times \P^1$. Using \eqref{eq:hodgenumbersZ}, the Hodge numbers of $Z = Z_{(\Diamond,\Diamond^\circ)}$ are found to be
\begin{equation}
 h^{1,1}(Z) = 2 \hspace{1cm} h^{2,1}(Z) = 33 \, ,
\end{equation}
which can easily be verified using the standard index and vanishing theorems. As detailed in Appendix \ref{sect:degcy3tobb}, this building block can also be found by degenerating a $K3$ fibred Calabi-Yau threefold. In particular, we may consider the Calabi-Yau hypersurface in $\P^3 \times \P^1$, which is given by a homogeneous polynomial of bidegree $(4,2)$. 

The lattice $N(Z_{(\Diamond,\Diamond^\circ)})$ is simply $(4)$ in this case (generated by the hyperplane class of $\P^3$) and the lattice $T$ is 
\begin{equation}
T = (-4) \oplus U^{\oplus 2} \oplus (-E_8)^{\oplus 2} \, .
\end{equation}
It follows that $K(Z_{(\Diamond,\Diamond^\circ)})=0$, which corresponds to the $K3$ fibration having no reducible fibres and hence no localized divisors. 

One may orthogonally glue two of these identical building blocks to a $G_2$ manifold. Here, the lattices $N_\pm = (4)$ are simply embedded into different $U$ summands of $\Gamma^{3,19}$. Note that this gluing is not only orthogonal, but also satisfies that $N_+ + N_-$ is already embedded into $\Gamma^{2,18}$. 
It follows that
\begin{equation}
\begin{aligned}
N_\pm \cap T_\mp & = (4) \\
N_+ \cap N_- & = 0 \\
T_+ \cap T_- & = (-4)^{\oplus 2} \oplus U \oplus (-E_8)^{\oplus 2} \\
|\Gamma^{3,19}/(N_+ + N_-)|& = 20 \\
|\Gamma^{3,19}/(N_\pm + T_\mp)|& = 1 \\
\Gamma^{3,19}/(T_+ + T_-) & = 0
\end{aligned}
\end{equation}
Evaluating \eqref{eq:bettiG2} we find
\begin{equation}
 b_2(J) = 0 \,,\hspace{.5cm} b_3(J) = 155 \,,\hspace{.5cm}b_4(J) = 155 \, ,
\end{equation}
which satisfies $b_2 + b_3 = 23 + 2\left(h^{2,1}(Z_+) + |K_+|  \right) + 2\left(h^{2,1}(Z_-) + |K_-|  \right)$.

Let us now consider the mirror $Z^\vee = Z_{(\Diamond^\circ,\Diamond)}$. From \eqref{eq:hodgenumbersZ} it follows that
\begin{equation}
 h^{1,1}(Z^\vee) = 53 \hspace{1cm} h^{2,1}(Z^\vee) = 0 \, .
\end{equation}
Furthermore, 
\begin{equation}
\begin{aligned}
N^\circ = N(Z^\vee) = \tilde{T}(Z) &=  (-4) \oplus U \oplus (-E_8)^{\oplus 2} \\
N(Z) = \tilde{T^\circ} = \tilde{T}(Z^\vee) &=  (4)
\end{aligned}
\end{equation}
so that the $|K(Z^\vee)| = 33$ by \eqref{eq:rkK} and we see \eqref{eq:dualrel2} at work. Using the mirror glueing as described in Section \ref{sect:mirrorgluing} we find for $Z^\vee$:
\begin{equation}
\begin{aligned}
N_\pm^\circ \cap T_\mp^\circ & = (4) \\
N_+^\circ \cap N_-^\circ & = (-4)^{\oplus 2} \oplus (-E_8)^{\oplus 2} \\
T_+^\circ \cap T_-^\circ & = U \\
|\Gamma^{3,19}/(N_+^\circ + N_-^\circ)|& = 2 \\
|\Gamma^{3,19}/(N_\pm^\circ + T_\mp^\circ)|& = 1 \\
|\Gamma^{3,19}/(T_+^\circ + T_-^\circ)| & = 18
\end{aligned}
\end{equation}
so that
\begin{equation}
 b_2(J^\vee) = 84 \,,\hspace{.5cm} b_3(J^\vee) =  71 \,,\hspace{.5cm}b_4(J^\vee) = 71 \, .
\end{equation}
Note that $b_2$ and $b_4$ are not swapped, but rather the $155$ classes in $H^4(J)$ are redistributed as $84+71$ for $J^\vee$. This is already familiar from the orbifold examples in \cite{Shatashvili:1994zw}.

As there is no torsion in $H^3(J)$, $H^4(J)$, $H^3(J^\vee)$ and $H^4(J^\vee)$ it follows that
\begin{equation}\label{eq:eqnotorsion}
\begin{aligned}
  &H^2(J,\Z) \oplus H^3(J,\Z) &=&\,\, H^2(J^\vee,\Z) \oplus H^3(J^\vee,\Z)  \\
=\,\,&H^2(J,\Z) \oplus H^4(J,\Z) &=&\,\, H^2(J^\vee,\Z) \oplus H^4(J^\vee,\Z)
\end{aligned}
\end{equation}
It is not hard to make similar examples which include torsion in $H^4(J,\Z)$ and $H^3(J^\vee,\Z)$. In all the examples we constructed, both the torsion in 
$H^2(J,\Z) \oplus H^4(J,\Z)$ (as expected from the general result \eqref{eq:equalityofh2h4}) and the torsion in $H^2(J,\Z) \oplus H^3(J,\Z)$ are preserved under the mirror map.

\subsection{Building Blocks fibred by an Elliptic $K3$ Surface}

We now consider examples of building blocks for which the fibre is an elliptic $K3$ surface. For a $K3$ fibration in Weierstrass form without degenerate $K3$ fibres the top $\Diamond_a^\circ$ and its dual $\Diamond_a$ have vertices
\begin{equation}
\Diamond^\circ_a = \left(\begin{array}{rrrrr}
-1 & 0 & 2 & 2 & 2 \\
0 & -1 & 3 & 3 & 3 \\
0 & 0 & -1 & 0 & 1 \\
0 & 0 & 0 & 1 & 0
\end{array}\right) \, ,\hspace{1cm} 
\Diamond_a = \left(\begin{array}{rrrrrr}
-2 & 1 & 1 & 1 & 1 & 1 \\
1 & 1 & 1 & 1 & 1 & -1 \\
0 & 6 & 6 & -6 & -6 & 0 \\
0 & 0 & -6 & 0 & -6 & 0
\end{array}\right)
\end{equation}
The Hodge numbers are
\begin{equation}
\begin{aligned}
h^{1,1}(Z_{(\Diamond_a,\Diamond_a^\circ)}) &= 3 \hspace{1cm} h^{2,1}(Z_{(\Diamond_a,\Diamond_a^\circ)}) &=& 112 \hspace{1cm} |N(Z_{(\Diamond_a,\Diamond_a^\circ)})| &=2\\
h^{1,1}(Z_{(\Diamond_a^\circ,\Diamond_a)}) &= 131 \hspace{1cm} h^{2,1}(Z_{(\Diamond_a^\circ,\Diamond_a)}) &=& 0 \hspace{1cm} |N(Z_{(\Diamond_a^\circ,\Diamond_a)})| &=18\\
\end{aligned}
\end{equation}
so that $K(Z_{(\Diamond_a,\Diamond_a^\circ)}) = 0$ and $|K(Z_{(\Diamond_a^\circ,\Diamond_a)})| = 131 -18 - 1 = 112$ as expected from \eqref{eq:dualrel2}. In particular, 
\begin{equation}
\begin{aligned}
 N(Z_{(\Diamond_a,\Diamond_a^\circ)}) & = U \\
 N(Z_{(\Diamond_a^\circ,\Diamond_a)}) & = U \oplus (-E_8)^{\oplus 2 }
\end{aligned}
\end{equation}
The $K3$ fibre of the mirror building block is hence also elliptically fibred with two $II^*$ fibres.

As a second example, let us consider a top for which every elliptic $K3$ fibre has a degenerate elliptic fibre of type $II^*$. The vertices of the corresponding top $\Diamond_b^\circ$ and its dual $\Diamond_b$ have vertices
\begin{equation}
 \Diamond_b^\circ = \left(\begin{array}{rrrrr}
-1 & 0 & 2 & 2 & 2 \\
0 & -1 & 3 & 3 & 3 \\
0 & 0 & -1 & 0 & 6 \\
0 & 0 & 0 & 1 & 0
\end{array}\right) \, , \hspace{1cm}
\Diamond_b = \left(\begin{array}{rrrrrr}
-2 & 1 & 1 & 1 & 1 & 1 \\
1 & 1 & 1 & 1 & 1 & -1 \\
0 & 6 & 6 & -1 & -1 & 0 \\
0 & 0 & -6 & 0 & -6 & 0
\end{array}\right)
\end{equation}
The Hodge numbers are
\begin{equation}
\begin{aligned}
h^{1,1}(Z_{(\Diamond_b,\Diamond_b^\circ)}) &= 17 \hspace{1cm} h^{2,1}(Z_{(\Diamond_b,\Diamond_b^\circ)}) &=& 66 \hspace{1cm} |N(Z_{(\Diamond_b,\Diamond_b^\circ)})| &= 10 \\
h^{1,1}(Z_{(\Diamond_b^\circ,\Diamond_b)}) &= 77 \hspace{1cm} h^{2,1}(Z_{(\Diamond_b^\circ,\Diamond_b)}) &=& 6 \hspace{1cm} |N(Z_{(\Diamond_b^\circ,\Diamond_b)})| &= 10 \\
\end{aligned}
\end{equation}
Now
\begin{equation}
\begin{aligned}
 N(Z_{(\Diamond_b,\Diamond_b^\circ)}) & = U \oplus (-E_8) \\
 N(Z_{(\Diamond_b^\circ,\Diamond_b)}) & = U \oplus (-E_8)
\end{aligned}
\end{equation}
It follows that $|K(Z_{(\Diamond_b,\Diamond_b^\circ)})| =  6 $ and $|K(Z_{(\Diamond_b^\circ,\Diamond_b)})| = 66$. Hence both building blocks have reducible $K3$ fibres. 

Let us now glue $Z_{(\Diamond_a,\Diamond_a^\circ)}$ with $Z_{(\Diamond_b,\Diamond_b^\circ)}$ by embedding the $N(Z_{(\Diamond_a,\Diamond_a^\circ)})$ and $N(Z_{(\Diamond_b,\Diamond_b^\circ)})$ orthogonal and perpendicular into $\Gamma^{3,19}$. We find 
\begin{equation}
\begin{aligned}
N_a \cap N_b & = 0 \\
N_a \cap T_b & = U \\
N_b \cap T_a & = U \oplus (-E_8) \\
T_a \cap T_b & = U \oplus (-E_8) \\
\Gamma^{3,19}/(N_a + N_b)& = U \oplus (-E_8) \\
\Gamma^{3,19}/(N_a + T_b)& = U \oplus (-E_8) \\
\Gamma^{3,19}/(N_b + T_a)& = U  \\
\Gamma^{3,19}/(T_a + T_b) & = 0  
\end{aligned}
\end{equation}
Hence 
\begin{equation}
 b_2(J) = 6 \,,\hspace{.5cm} b_3(J) = b_4(J) = 385 \, ,
\end{equation}

For the mirror $G_2$ manifold $J^\vee$ we now glue $Z_{(\Diamond_a^\circ,\Diamond_a)}$ with $Z_{(\Diamond_b^\circ,\Diamond_b)}$ using the same embedding as above with the replacement $N = \tilde{T^\circ}$ and $N^\circ = \tilde{T}$. Now
\begin{equation}
\begin{aligned}
N_a^\circ \cap N_b^\circ & = (-E_8) \\
N_a^\circ \cap T_b^\circ & =  U \oplus (-E_8) \\
N_b^\circ \cap T_a^\circ & = U \\
T_a^\circ \cap T_b^\circ & = U \\
\Gamma^{3,19}/(N_a^\circ + N_b^\circ)& = U \\
\Gamma^{3,19}/(N_a^\circ + T_b^\circ)& = U \\
\Gamma^{3,19}/(N_b^\circ + T_a^\circ)& = U \oplus (-E_8) \\
\Gamma^{3,19}/(T_a^\circ + T_b^\circ) & = (-E_8)
\end{aligned}
\end{equation}
Consequently,
\begin{equation}
 b_2(J^\vee) = 186 \,,\hspace{.5cm} b_3(J^\vee) = b_4(J^\vee) = 205 \, .
\end{equation}
so that we find again that \eqref{eq:eqnotorsion} holds. Again, the Betti numbers $b_2$ and $b_4$ are not swapped but rather redistributed. As we have used orthogonal gluing again, the Betti numbers of $J$ and $J^\vee$ satisfy \eqref{eq:b2pb3fororthgluing} also in this examples.

Starting from this example, it is easy to describe singular transitions on the level of the building blocks in which the polarizing lattice of the $K3$ fibre changes, e.g. by blowing down components of the $II^*$ fibres (with a subsequent deformation) or colliding singular elliptic fibres of the $K3$ surfaces (followed by a resolution). Of course, we can also have transitions in which the lattices $K$ change by colliding singular $K3$ fibres (followed by a blowup) or blowing down components of the reducible $K3$ fibres (followed by a deformation of the building block). As is familiar from the case of reflexive polytopes, such transitions can be efficiently described using the dual pairs of tops. Furthermore, given our mirror construction, we can track the behaviour of the glued $G_2$ manifold as well as its mirror when we perform such changes. Even though, it is still an open question if the singular manifolds in the middle of the transition allow metrics of $G_2$ holonomy, using this technique for the example discussed above allows to construct a plethora of closely related smooth mirror pairs.

\section*{Acknowledgements}

We thank Chris Beem, Ruben Minasian, David R. Morrison, Sakura Schafer-Nameki and Alessandro Tomasiello for discussions. Special thanks go to Samson Shatashvili for bringing his 1997 paper with Cumrun Vafa to our attention during the workshop on Special Holonomy held at the Simons Center for Geometry and Physics in October 2016. AB thanks the Simons Center for Geometry and Physics for hospitality during the workshop on Special Holonomy where this project was conceived. MDZ thanks the Institute of Mathematics of the University of Oxford and the department of Physics of the University ``Bicocca'' of Milan for the kind hospitality during the completion of this work. The research of AB is supported by the STFC grant ST/L000474/1 and the EPSCR grant EP/J010790/1.

\appendix

\section{Toolkit for computing $H^\bullet(J,\mathbb{Z})$ from building blocks}\label{FORMULAHOM}

The diffeomorphism $\Xi$ we discussed around Eqn.\eqref{THADIFFEOH}, in particular induces an identification $g^*:\Sigma_\pm \leftrightarrow \Sigma_\pm$ between the three-planes $\Sigma_\pm$ determining a point in the Teichm\"uller space of Ricci-flat metrics. Conversely, there is a unique diffeomorphism for each lattice isometry $g_\Gamma^{3,19} : H^2(S_{0+},\mathbb{Z}) \rightarrow  H^2(S_{0-},\mathbb{Z})$ inducing $g^*:\Sigma_\pm \leftrightarrow \Sigma_\pm$ by the global Torelli theorem.
We may choose markings $h_\pm: \Gamma^{3,19}  \cong H^2(S_{0,\pm},\mathbb{Z})$ on the K3 surfaces such that the condition in Eqn.\eqref{HKAROTAH} simply becomes 
\be\label{eq:hkrot2}
\begin{aligned}
\omega_{S_0 \pm} &= Re(\Omega_{S_{0\mp}}) \\
Im(\Omega_{S_{0\pm}}) &= - Im(\Omega_{S_{0\mp}}) \, ,
\end{aligned}
\ee
This marking defines primitive embeddings $N_\pm \hookrightarrow \Gamma^{3,19}$. Let us denote the orthogonal complements of $N_\pm$ in $\Gamma^{3,19} $ by $T_\pm$. The integral cohomology groups of the resulting $G_2$ manifolds $J$ are then given by \cite{Corti:2012kd}:
\begin{equation}\label{eq:bettiG2}
\begin{aligned}
H^1(J,\mathbb{Z}) & =   0 \\
H^2(J,\mathbb{Z}) & =  N_+ \cap N_- \oplus K_+ \oplus K_- \\
H^3(J,\mathbb{Z}) & = \mathbb{Z}[S] \oplus \Gamma^{3,19} /(N_+ + N_-) \oplus (N_- \cap T_+) \oplus (N_+ \cap T_-)\\
& \hspace{1cm} \oplus H^3(Z_+)\oplus H^3(Z_-) \oplus K_+ \oplus K_- \\
H^4(J,\mathbb{Z}) & = H^4(S) \oplus (T_+ \cap T_-) \oplus \ \Gamma^{3,19} /(N_- + T_+) \oplus \Gamma^{3,19} /(N_+ + T_-) \\
& \hspace{1cm} \oplus  H^3(Z_+)\oplus H^3(Z_-) \oplus K_+^* \oplus K_-^* 
\end{aligned}
\end{equation}
Here, the group $K$ is defined as
\begin{equation}
K \equiv \mbox{ker} (\rho)/[S_0] \, .
 \end{equation}
and $K^*$ is its dual.

There is a particularly simple class of glueings which are called `orthogonal' in \cite{Corti:2012kd}: here $N_+\otimes \R$ and $N_-\otimes \R$ are embedded orthogonally (but not necessarily perpendicular) into $\Gamma^{3,19} $. For these, the primitive embeddings $N_\pm \hookrightarrow \Gamma^{3,19} $ are such that
\begin{equation}\label{eq:orthgluecond}
N_\pm \otimes \R = (N_\pm\otimes\R)\cap(N_\mp\otimes\R) \oplus (N_\pm\otimes\R \cap T_\mp\otimes\R) \, .
\end{equation}
As a consequence, the dimension of 
\begin{equation}
 \left(N_+ \cap N_-\right) \oplus \Gamma^{3,19} /(N_+ + N_-) \oplus (N_- \cap T_+) \oplus (N_+ \cap T_-)
\end{equation}
is always equal to the dimension of the lattice $\Gamma^{3,19} $, $|\Gamma^{3,19} | = 22$, so that we find 
\begin{equation}\label{eq:b2pb3fororthgluing}
b_2(J) + b_3(J) = 23 + 2\left(h^{2,1}(Z_+) + |K_+|  \right) + 2\left(h^{2,1}(Z_-) + |K_-|  \right) \, ,
\end{equation}
as a consequence of \eqref{eq:bettiG2}.

\section{Hodge numbers, $N$ and $K$ for a building block from dual tops}\label{HpqNKs}

In this appendix we list some results from \cite{Braun:2016igl} about topological properties of building blocks constructed from tops as in Section \ref{sect:tops_as_build}.

The Hodge numbers of a building block constructed from a pair of dual tops $(\Diamond,\Diamond^\circ)$ are:
\begin{equation}
\begin{aligned}\label{eq:hodgenumbersZ}
h^{1,1}(Z_{(\Diamond,\Diamond^\circ)}) &= h^{2,2}(Z) = -4 + \sum_{\Theta^{[3]}\in \Diamond} 1 + \sum_{\Theta^{[2]}\in \Diamond} \ell^*(\sigma_n(\Theta^{[2]})) + \sum_{\Theta^{[1]}\in \Diamond}(\ell^*(\Theta^{[1]})+1)\cdot \ell^*(\sigma_n(\Theta^{[1]}))  \\
h^{2,1}(Z_{(\Diamond,\Diamond^\circ)}) &= \ell(\Diamond) - \ell(\Delta_F) + \sum_{\Theta^{[2]} < \Diamond} \ell^*(\Theta^{[2]})\cdot \ell^*(\sigma_n(\Theta^{[2]})) - \sum_{\Theta^{[3]} < \Diamond}\ell^*(\Theta^{[3]})  \\
h^{3,0}(Z_{(\Diamond,\Diamond^\circ)}) &= \ell^*(\Diamond)=0
\end{aligned} 
\end{equation}
Here $\Theta^{[k]}$ denotes a $k$-dimensional face of the ${\bf M}$-lattice top $\Diamond$, $\ell(\Theta^{[k]})$ counts the number of integral points on such a face and $\ell^*(\Theta^{[k]})$ the number of lattice points in the relative interior of such a face. $\sigma_n(\Theta^{[k]})$ is the cone in the normal fan of $\Diamond$ associated with $\Theta^{[k]}$ and $\ell^*(\sigma_n(\Theta^{[k]}))$ counts the number of integral points of $\Diamond^\circ \cup \nu_0$, i.e. the number of rays of $\Sigma$, in the relative interior of this cone. 

A one-dimensional face $\Theta_F^{\circ [1]}$ is called non-vertically embedded (nve) if there is no face $\Theta^{\circ [2]}$ of $\Diamond^\circ$ perpendicular to $F$ which contains $\Theta_F^{\circ [1]}$ in its boundary, and it is called vertically embedded (ve) otherwise. As shown in \cite{Braun:2016igl} a pair of dual faces $\Theta_F^{\circ [1]}, \Theta_F^{[1]}$, under the polar duality of $(\Delta_F,\Delta_F^\circ)$, is always sitting in $\Diamond^\circ, \Diamond$ such that one of them is ve, and the other one is nve. Whenever $\ell^*(\Theta^{\circ [1]}_F) \cdot \ell^*(\Theta^{[1]}_F)$ is non-zero, there are divisors $D_i$ of the toric ambient space $\P_\Sigma$ which split into several disjoint irreducible components $D_i^\alpha$ on $X_{(\Delta_F,\Delta^\circ_F)}$. These are associated with lattice points interior to $\Theta^{\circ [1]}_F$ and the number of irreducible components is given by $\ell^*(\Theta^{[1]}_F)+1$, where $\Theta^{\circ [1]}_F$ and $\Theta^{[1]}_F$ are dual faces on $\Delta_F,\Delta^\circ_F$.
This is not necessarily the case on $Z_{(\Diamond,\Diamond^\circ)}$, and the individual $D_i^\alpha$ are only contained in $N$ if the face $\Theta_F^{\circ [1]}$ (which is also a face of $\Diamond^\circ$) is nve in $\Diamond^\circ$ \cite{Braun:2016igl}. Consequently, the lattice $N$ is given by
\begin{equation}\label{eq:latticeN}
 N(Z_{(\Diamond,\Diamond^\circ)}) = \mbox{Pic}_{tor}(X_{\Delta_F,\Delta^\circ_F}) + \sum_{\mbox{nve} \,\Theta_{F}^{\circ[1]}} L(\Theta_{F}^{\circ[1]},\Theta_F^{[1]})\, .
\end{equation}
Here the lattice $\mbox{Pic}_{tor}(X_{\Delta_F,\Delta^\circ_F})$ is the lattice of cycles obtained by restricting toric divisors of the ambient toric space of the fibre (this does not depend on a triangulation of $\Delta^\circ_F$) and the lattice $L(\Theta_{F}^{\circ[1]},\Theta_F^{[1]})$ contains all of the irreducible components $D_i^\alpha$ of divisors $D_i$ of the ambient space which become reducible on $X_{(\Delta_F,\Delta^\circ_F)}$.

The rank of $N$ is 
\begin{equation}
|N(Z_{(\Diamond,\Diamond^\circ)})| = |\mbox{Pic}_{tor}(X_{\Delta_F,\Delta^\circ_F})| + \sum_{\mbox{nve} \,\Theta_{F}^{\circ[1]}}  \ell^*(\Theta_F^{\circ [1]})\ell^*(\Theta_F^{[1]}) \, ,
\end{equation}
where $|\mbox{Pic}_{tor}(X_{\Delta_F,\Delta^\circ_F})|= \ell^1(\Delta_F^\circ) -3$. It follows that the rank of $K(Z_{(\Diamond,\Diamond^\circ)})$ is
\begin{equation}\label{eq:rkK}
|K(Z_{(\Diamond,\Diamond^\circ)})| =h^{1,1}(Z_{(\Diamond,\Diamond^\circ)})- |N(Z_{(\Diamond,\Diamond^\circ)})|-1 \, . 
\end{equation}
The divisors contributing to $K(Z_{(\Diamond,\Diamond^\circ)})$ correspond to singular fibre components, which in turn correspond to lattice points on $\Diamond$ above $F$ as well as points interior to two-dimensional faces of $\Delta^\circ_F$.

\section{Topological Properties of Mirror Building Blocks}\label{sect:mirrorbuildblocksdetails}

In this appendix, we prove the two key properties for a pair of mirror building blocks $Z_{(\Diamond,\Diamond^\circ)}$ and $Z_{(\Diamond^\circ,\Diamond)}$ stated in the beginning of Section \ref{sect:mirror}:
\begin{enumerate}
 \item[a)] The lattices $N(Z_{(\Diamond,\Diamond^\circ)})$ and $N(Z_{(\Diamond^\circ,\Diamond)})$ admit a primitive embedding
 \begin{equation}\label{eq:dualrel1ap}
N(Z_{(\Diamond,\Diamond^\circ)}) \oplus N(Z_{(\Diamond^\circ,\Diamond)}) \oplus U \hookrightarrow H^{2}(S,\mathbb{Z}) \, .
\end{equation}

\item[b)] For a mirror pair of building blocks, the rank of the lattice $K$ and the Hodge number $h^{2,1}$ satisfy
\begin{equation}\label{eq:dualrel2ap}
\begin{aligned}
|K(Z_{(\Diamond,\Diamond^\circ)})| = h^{2,1}(Z_{(\Diamond^\circ,\Diamond)}) \\
|K(Z_{(\Diamond^\circ,\Diamond)})| = h^{2,1}(Z_{(\Diamond,\Diamond^\circ)}) 
\end{aligned} 
\end{equation}
\end{enumerate}

Let us start with relation $a)$. For a pair of reflexive three-dimensional polytopes $\Delta_F,\Delta^\circ_F$, the $K3$ surface $X_{(\Delta_F,\Delta_F^\circ)}$ is lattice polarized by a lattice 
\begin{equation}\label{eq:pick3}
n_{(\Delta_F,\Delta^\circ_F)} = \mbox{Pic}_{tor}(X_{\Delta_F^\circ,\Delta_F}) + \sum_{\Theta^{\circ [1]}} L(\Theta^{\circ [1]},\Theta^{[1]})
\end{equation}
of dimension
\begin{equation}
|n_{(\Delta_F,\Delta^\circ_F)}| = \ell^1(\Delta_F^\circ) - 3 +\sum_{(\Theta^{[1]},\Theta^{\circ [1]})} \ell^*(\Theta^{[1]})\ell^*(\Theta^{\circ [1]}) \, ,
\end{equation}
where $\ell^1(\Delta_F)$ counts the number of lattice points on the one-skeleton of $\Delta_F^\circ$. Here, $ \ell^1(\Delta_F^\circ)-3$ counts the dimension of independent divisor classes in $\mbox{Pic}_{tor}(X_{\Delta_F,\Delta_F^\circ})$, which are restrictions of toric divisors of the ambient space, and the correction term takes into account the fact that toric divisors may become reducible on the $K3$ hypersurface. By a straightforward evaluation of this formula for all $4319$ cases, the complete list reflexive three-dimensional polytopes classified by \cite{Kreuzer:1998vb},
it can be shown that 
\begin{equation}
|n_{(\Delta_F,\Delta^\circ_F)}| + |n_{(\Delta_F^\circ,\Delta_F)}| =  20 +\sum_{(\Theta^{[1]},\Theta^{\circ [1]})} \ell^*(\Theta^{[1]})\ell^*(\Theta^{\circ [1]})  
\end{equation}
Hence we cannot simply associate these two lattice polarized families as mirror pairs of $K3$ surfaces, as this would imply the absence of the correction term on the right-hand side. The intuitive interpretation of this result is that this correction term accounts both for K\"ahler deformations associated with non-toric divisors and non-polynomial complex structure deformations (of the mirror). Consequently, one may conjecture that $n_{(\Delta_F,\Delta^\circ_F)}$ and $\mbox{Pic}_{tor}(X_{\Delta_F^\circ,\Delta_F})$ have a primitive embedding 
\begin{equation}
\mbox{Pic}_{tor}(X_{\Delta_F^\circ,\Delta_F}) \oplus n_{(\Delta_F,\Delta^\circ_F)} \oplus U \hookrightarrow \Gamma^{3,19}  \, ,
\end{equation}
corresponding to a mirror family of lattice polarized $K3$ surfaces. This relations was shown to be true by \cite{Rohsiepe:2004st} by computing the discriminant forms for all $4319$ cases. 

In the present case, the result of \cite{Rohsiepe:2004st} implies the existence of the primitive embedding \eqref{eq:dualrel1ap} in the case that 
\begin{equation}
\begin{aligned}
&& N(Z_{(\Diamond,\Diamond^\circ)}) & = n_{(\Delta_F,\Delta^\circ_F)} \hspace{.3cm} \rightarrow \hspace{.3cm} N(Z_{(\Diamond^\circ,\Diamond)}) & =\mbox{Pic}_{tor}(X_{\Delta_F^\circ,\Delta_F}) \\
\mbox{or} && N(Z_{(\Diamond^\circ,\Diamond)}) & = n_{(\Delta_F^\circ,\Delta_F)} \hspace{.3cm} \rightarrow \hspace{.3cm} N(Z_{(\Diamond,\Diamond^\circ)}) & =\mbox{Pic}_{tor}(X_{\Delta_F,\Delta_F^\circ})  
\end{aligned}
\end{equation}
which means that the one-dimensional faces of $\Delta_F^\circ$ for which $L(\Theta^{\circ [1]}_F,\Theta^{[1]}_F)$ is non-trivial are either all ve or all nve. In this case the non-trivial $L(\Theta^{\circ [1]}_F,\Theta^{[1]}_F)$ are all associated with the lattice polarization of $X_{(\Delta_F,\Delta_F^\circ)}$, or they are all associated with $X_{(\Delta_F^\circ,\Delta_F)}$. 

For more general tops, some of the $L(\Theta^{\circ [1]}_F,\Theta^{[1]}_F)$ will contribute to $N(Z_{(\Diamond,\Diamond^\circ)})$, whereas others will contribute to $N(Z_{(\Diamond^\circ,\Diamond)})$ and we need a more general result. First note each of the summands in the correction factor $\ell^*(\Theta^{[1]})\ell^*(\Theta^{\circ [1]})$ will either contribute to $N(Z_{(\Diamond,\Diamond^\circ)})$ or $N(Z_{(\Diamond^\circ,\Diamond)})$, so that the dimensions work out. However, as $L(\Theta^{\circ [1]}_F,\Theta^{[1]})_F \neq L(\Theta^{[1]}_F,\Theta^{\circ [1]}_F)$ (see \cite{Braun:2016igl} for a detailed description of these lattices) and furthermore $L(\Theta^{\circ [1]}_F,\Theta^{[1]}_F) \cap \mbox{Pic}_{tor}(X_{\Delta_F^\circ,\Delta_F}) \neq 0$, \eqref{eq:dualrel1ap} is still a very non-trivial result. We have proven \eqref{eq:dualrel1ap}
by computing the discriminant forms for each possible such pair of lattices and checking that they satisfy \cite{MR525944}
\begin{equation}\label{eq:disrforms}
\begin{aligned}
 G(N_{(\Diamond,\Diamond^\circ)}) &\cong G(N_{(\Diamond^\circ,\Diamond)}) \\ 
 q(N_{(\Diamond,\Diamond^\circ)}) &\cong - q(N_{(\Diamond^\circ,\Diamond)}) \, .
\end{aligned}
\end{equation}
This is possible as there are only finitely many reflexive polytopes and for each pair of polytopes there are only finitely many options for which one-dimensional faces are vertically embedded or non-vertically embedded, i.e. contribute to $N(Z_{(\Diamond,\Diamond^\circ)})$ or $N(Z_{(\Diamond^\circ,\Diamond)})$. Extending the work of \cite{Rohsiepe:2004st}, the present authors checked \eqref{eq:disrforms} for all cases using the computer algebra system Sage \cite{sage}.

Let us now prove relation $b)$, for which we need to evaluate the formulas \eqref{eq:rkK} and \eqref{eq:hodgenumbersZ}. As a preparation, let us quote a central result of \cite{Braun:2016igl} about the normal fan of tops: \\
The normal fan $\Sigma_n(\Diamond)$ of a top $\Diamond$ is equal to the face fan $\Sigma_f(\Diamond^\circ)$ of $\Diamond^\circ$ except for vertically embedded faces $\Theta_F^{\circ [k]}$ and the faces $\Theta_{F,+}^{\circ [k+1]}$ and $\Theta_{F,\nu_0}^{\circ [k+1]}$ which are connected to them above and below $F$. For such faces, the normal fan $\Sigma_n(\Diamond)$ contains only a single $k+2$-dimensional cone which is the union of $\sigma_f(\Theta_{F,+}^{\circ [k+1]})$ and $\sigma_f(\Theta_{F,\nu_0}^{\circ [k+1]})$, where $\sigma_f(\Theta^\circ)$ denotes the cone over the face $\Theta$. Consequently, $\sigma_f(\Theta_F^{\circ [k]})$ is not present in $\Sigma_n(\Diamond)$ for vertically embedded faces $\Theta_F^{\circ [k]}$.\\
Here, a vertically embedded face $\Theta_F^{\circ [k]}$ is a face of $\Diamond^\circ$ sitting in $F$ (so it is also a face of $\Delta_F^\circ$) which is contained in a vertical (i.e. perpendicular to $F$) face of $\Diamond^\circ$. In this case its dual face $\Theta^{[2-k]}$ under polar duality of $(\Delta_F, \Delta_F^\circ)$ is non-vertically embedded, i.e. it does not sit in a vertical face of $\Diamond$. We will use $\Theta_{F,V}^{[k]}$ to denote vertically embedded faces on $\Delta_F$, $\Theta_{F,NV}^{[k]}$ to denote non-vertically embedded faces and $\Theta_{R}^{[k]}$ for faces not contained in $F$. Furthermore, vertical faces are denoted by $\Theta_V$ and non-vertical faces by $\Theta_{NV}$.

After this preparation, let us start with the formula for $h^{2,1}(Z_{(\Diamond, \Diamond^\circ)})$, \eqref{eq:hodgenumbersZ}.
\begin{equation}
 \begin{aligned}
h^{2,1}(Z_{(\Diamond,\Diamond^\circ)}) & = \ell(\Diamond) - \ell(\Delta_F) + \sum_{\Theta^{[2]} < \Diamond} \ell^*(\Theta^{[2]})\cdot \ell^*(\sigma_n(\Theta^{[2]})) - \sum_{\Theta^{[3]} < \Diamond}\ell^*(\Theta^{[3]})  \\
& = - \ell(\Delta_F) + \ell^0(\Diamond) + \ell^1(\Diamond)  + \sum_{\Theta_R^{[2]}} \ell^*(\Theta_R^{[2]})\cdot (1+\ell^*(\Theta_R^{[1]})) \\
 & \hspace{3ex}+ \sum_{\Theta_{F,V}^{[2]}} \ell^*(\Theta_{FV}^{[2]}) + \sum_{\Theta_{F,NV}^{[2]}} \ell^*(\Theta_{F,NV}^{[2]})\cdot \left(1 + \ell^*(\Theta_V^{\circ[1]})) 
+ \ell^*(\Theta_{F,V}^{\circ[0]})\right) 
\end{aligned}
\end{equation}
Here $\Theta_R^{[2]}$ and $\Theta_R^{[1]}$ are dual faces for which $\sigma_f(\Theta_R^{[1]}) = \sigma_n(\Theta_R^{[2]})$ and $(\Theta_{F,NV}^{[2]},\Theta_{F,V}^{\circ[0]})$ are dual faces under polar duality of $(\Delta_F, \Delta_F^\circ)$. Finally, $\Theta_V^{\circ[1]}$ is the vertical face bounded by $\Theta_{F,V}^{\circ[0]}$. Hence
\begin{equation}
 \begin{aligned}
h^{2,1}(Z_{(\Diamond,\Diamond^\circ)}) & =   - \ell(\Delta_F) + \ell^0(\Diamond) + \ell^1(\Diamond) + \sum_{\Theta^{[2]} \neq \Theta_{F,V}^{[2]}} \ell^*(\Theta^{[2]})
\cdot (1 + \ell^*(\Theta^{\circ [1]})) \\
& \hspace{3ex} + \sum_{\Theta_{F,V}^{[2]}} \ell^*(\Theta_{F,V}^{[2]}) + \sum_{\Theta_{F,NV}^{[2]}} \ell^*(\Theta_{F,NV}^{[2]}) 
 \end{aligned}
\end{equation}
Using the fact that the last line is equal to $\ell(\Delta_F) - \ell^1(\Delta_F) -1$, we finally find
\begin{equation}
h^{2,1}(Z_{(\Diamond,\Diamond^\circ)}) =   \ell^0(\Diamond) + \ell^1(\Diamond) + \sum_{\Theta^{[2]} \neq \Theta_{F,V}^{[2]}} \ell^*(\Theta^{[2]})
\cdot (1 + \ell^*(\Theta^{\circ [1]})) - \ell^1(\Delta_F) - 1
\end{equation}

Let us now evaluate $|K(Z_{(\Diamond^\circ,\Diamond)})|$:
\begin{equation}
\begin{aligned}
|K(Z_{(\Diamond^\circ,\Diamond)})| & = -4 + \sum_{\Theta^{\circ [3]}} 1 + \sum_{\Theta^{\circ [2]}} \ell^*(\sigma_n(\Theta^{\circ [2]}))  
+ \sum_{\Theta^{\circ [1]}} (\ell^*(\Theta^{\circ [1]})+1)\cdot \ell^*(\sigma_n(\Theta^{\circ [1]})) \\
& \hspace{3ex} -(\ell^1(\Delta_F) -3) - \sum_{\Theta_{F,V} \subset \sigma_n(\Theta^{\circ [1]})} \ell^*(\Theta^{\circ [1]}) \cdot \ell^*(\Theta^{[1]}) -1 \, .
\end{aligned}
\end{equation}
Using the dual faces on $\Diamond$ (where they exist) we find 
\begin{equation}
\begin{aligned}
|K(Z_{(\Diamond^\circ,\Diamond)})| & = -1 + \sum_{\Theta^{[0]}} 1 - \sum_{\Theta_{F,V}^{[0]}} 1 \\
& \hspace{3ex} + \sum_{\Theta^{[1]}} \ell^*(\Theta^{[1]}) + \sum_{\Theta_{F,V}^{[0]}} 1 - \sum_{\Theta_{F,V}^{[1]}} \ell^*(\Theta_{F,V}^{[1]}) \\
& \hspace{3ex}+ \sum_{\Theta^{[2]} \neq \Theta_{F,V}^{[2]}} \ell^*(\Theta^{[2]})\cdot (\ell^*(\Theta^{\circ [1]})+1) \\
& \hspace{3ex}+ \sum_{\Theta_{F,V}^{[1]}\subset \sigma_n(\Theta^{[\circ 1]})} \ell^*(\Theta_{F,V}^{[1]}) \cdot (1 + \ell^*(\Theta^{[\circ 1]})) \\
& \hspace{3ex}- \sum_{(\Theta_{F,V}^{[1]},\Theta_{F,NV}^{[1]})} \ell^*(\Theta_{F,V}^{[1]})\cdot \ell^*(\Theta_{F,NV}^{\circ [1]}) - \ell^1(\Delta_F) \\
& = \ell^0(\Diamond) + \ell^1(\Diamond) + \sum_{\Theta^{[2]} \neq \Theta_{F,V}^{[2]}} \ell^*(\Theta^{[2]})
\cdot (1 + \ell^*(\Theta^{\circ [1]})) - \ell^1(\Delta_F) - 1
\end{aligned}
\end{equation}
which agrees with the expression for $h^{2,1}(Z_{(\Diamond,\Diamond^\circ)})$ computed above ! Here, the $-1$ in the first line is due to the face
$\Theta_0^{[3]}=\Diamond \cap F$, which does not correspond to a face of $\Diamond^\circ$.

\section{Building Blocks and Degenerations of $K3$ fibred Calabi-Yau Threefolds}\label{sect:degcy3tobb}

Given two projecting tops (see Section \ref{sect:tops_as_build} for definitions) $\Diamond^\circ_a$ and $\Diamond^\circ_b$ which share the same $\Delta_F$, we may form a reflexive polytope $\Delta^\circ = \Diamond^\circ_a + \Diamond^\circ_b$. Here, we of course have to let one of the two, say $\Diamond^\circ_a$ to be above the plane $F$ (defined by $\Delta_F$) and the other, say $\Diamond^\circ_b$, below this plane. In this appendix, the '$+$' sign indicates that we simply take the (convex hull of) the union of the summands.

In this Section we demonstrate how a toric Calabi-Yau hypersurface $X_{(\Delta,\Delta^\circ)}$, can degenerate into the building blocks associated with the tops $\Diamond^\circ_a$ and $\Diamond^\circ_b$, $Z_{(\Diamond^\circ_a,\Diamond_a)}$ and $Z_{(\Diamond^\circ_b,\Diamond_b)}$. This limit can be thought of as a generalization of the degeneration of an elliptic $K3$ surface into two rational elliptic surfaces. 
As we also have that $\Delta = \Diamond_a + \Diamond_b$ for the polar dual reflexive polytope $\Delta$ of $\Delta^\circ$ and the dual tops (as defined in Section \ref{sect:tops_as_build}), the mirror $X_{(\Delta,\Delta^\circ)}$ has a similar degeneration limit into the mirror building blocks $Z_{(\Diamond_a,\Diamond^\circ_a)}$ and $Z_{(\Diamond_b,\Diamond^\circ_b)}$.

\begin{figure}[h!]
\begin{center}
  \includegraphics[height=3cm]{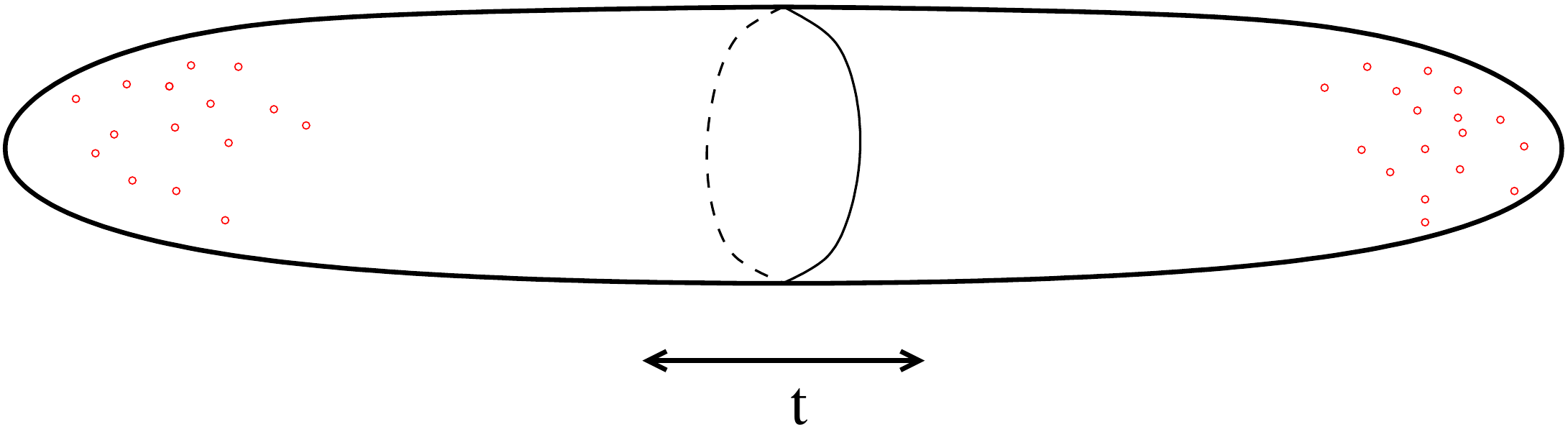}
 \caption{\label{fig:cydeg} In the degeneration limit of a $K3$ fibred Calabi-Yau threefold $X_{(\Delta,\Delta^\circ)}$, the singular fibres are localized towards the two ends of the elongated base. Cutting along the $S^1$ in the middle, we obtain the (open versions of) the building blocks $Z_{(\Diamond_a,\Diamond^\circ_a)}$ and $Z_{(\Diamond_b,\Diamond^\circ_b)}$. }
\end{center}
\end{figure}

In the light of the SYZ fibration, we may think of these two degeneration limits as follows. In the limit, the $\P^1$ base of $X_{(\Delta^\circ,\Delta)}$, which is of course $K3$ fibred, becomes infinitely long and effectively starts to look like a cylinder $S^1 \times \{t\}$. The singular fibres of the $K3$ fibration move to the two ends of this cylinder and the fibre becomes constant (equal to $S_0$) in the middle of this cylinder. In this picture, the open version of the building block (which is what is glued in the TCS construction) are found by cutting $X_{(\Delta,\Delta^\circ)}$ in two halves in the middle of the cylinder at $t=0$. Whereas it becomes non-trivial towards the ends of the interval, the SYZ fibration is very simple in the middle: it is composed of the SYZ fibration of the $K3$ surface $S_0$ and the $S^1$ of the cylinder. If we apply mirror symmetry, i.e. three T-dualities, in this limit, we hence end up again with a Calabi-Yau threefold of the same structure, but now with the mirror K3 surface as the constant fibre in the middle of the interval. As mirror symmetry is realized by swapping the roles of $\Delta$ and $\Delta^\circ$ for $X_{(\Delta^\circ,\Delta)}$, it must be that performing three T-dualities along the SYZ fibres turns $Z_{(\Diamond_a,\Diamond^\circ_a)}$ into $Z_{(\Diamond^\circ_a,\Diamond_a)}$.

Such a limit can be defined as follows: one first introduces a specific one-parameter family $\mathcal{X}_\zeta$ of threefolds $X_{(\Delta,\Delta^\circ)}$ parametrized by a coordinate $\zeta_a$, such that the fibre at $\zeta_a = 0$ is singular. After an appropriate blow-up, the family becomes smooth and the central fibre, which is now given by $\zeta_a \zeta_b = 0$ becomes reducible. These two components are nothing but the two building blocks, which can hence be found by setting $\zeta_a = 0$ and $\zeta_b=0$ in the family $\mathcal{X}_\zeta$. Taking inspiration from \cite{1307.6514, Cvetic:2015uwu}, we will describe this whole set-up by introducing a toric ambient space and defining equation for the whole family after the blow-up. Let us first describe the set-up in detail. Let us assume that we are given two projecting tops $\Diamond^\circ_a$ and $\Diamond^\circ_b$ which intersect along $\Delta^\circ_F$. We can chose $\Delta^\circ_F$ to be embedded in $\R^4$ such that its vertices have the form $\nu = (\nu_3,0)$. Furthermore, $\Diamond^\circ_a$ is assumed to be above $F$, i.e. its vertices have the form $\nu = (\nu_3,n_+)$ with the last coordinate $n_+ \geq 0 $, whereas $\Diamond^\circ_b$ is below $F$, so that its vertices have the form $\nu = (\nu_3,n_-)$ with the last coordinate $n_- \leq 0 $. Taking the union of two such tops we obtain a reflexive polytope $\Delta^\circ$. The polar dual $\Delta$ is then formed from $\Diamond_a$ and $\Diamond_b$ which intersect in $\Delta_F$. As follows from their definition, $\Diamond_a$ is now below $F$ and $\Diamond_b$ is above $F$. 

The threefold $X_{(\Delta^\circ,\Delta)}$ is constructed by fixing a hypersurface equation in an ambient space $\mathbb{P}_\Sigma$ obtained from a triangulation of $\Delta^\circ$ compatible with the $K3$ fibration. To embed the family $\mathcal{X}_\zeta$ we extend this ambient space to $\mathbb{P}_\Sigma^\zeta$ by introducing two new rays 
\begin{equation}
 \nu_{\zeta_a} = (0,0,0,1,1)\,\, , \hspace{2cm} \nu_{\zeta_b} = (0,0,0,0,1) \, .
\end{equation}
Let us denote the polytope formed by taking the convex hull of $\Delta^\circ$ together with $\nu_{\zeta_a}$ and $\nu_{\zeta_b}$ by $\Delta^\circ_\zeta$. 
The maximal cones of the fan of $\mathbb{P}_\Sigma^\zeta$ are found by taking
\begin{equation}
 \langle \sigma_a, \nu_{\zeta_a}\rangle \,\, , \hspace{1cm}  \langle \sigma_b, \nu_{\zeta_b}\rangle \,\, , \hspace{1cm}  \langle \sigma_F, \nu_{\zeta_a},\nu_{\zeta_b}\rangle \, ,
\end{equation}
where $\sigma_a$, $\sigma_b$ and $\sigma_F$ are cones of $\mathbb{P}_\Sigma$ ending on maximal dimensional faces of $\Diamond^\circ_a$, $\Diamond^\circ_b$, and $\Delta^\circ_F$, respectively. Note that this means that the $SR$-ideal of $\mathbb{P}_\Sigma^\zeta$ is such that $\zeta_a$ cannot vanish simultaneously with any of the coordinates associate with points on $\Diamond^\circ_b$ below $F$ and $\zeta_b$ cannot vanish simultaneously with any of the coordinates associate with points on $\Diamond^\circ_a$ above $F$ 

We now define the family $\mathcal{X}_\zeta$ as given by
\begin{equation}\label{eq:familyXdeglimit}
P_\zeta = \sum_{m \in \Delta_\zeta} c_m \prod_{\nu \in \Delta^\circ_\zeta} z_\nu^{\langle m,\nu\rangle +1} = 0  
\end{equation}
in  $\mathbb{P}_\Sigma^\zeta$. Here each monomial corresponds to a point on the polytope
\begin{equation}
 \Delta_\zeta = \left(\begin{array}{c}
                  \Diamond_b \\ 0
                 \end{array}\right)
+
\left(\begin{array}{c}
                  \Diamond_b \\ 1
                 \end{array}\right)
+
\left(\begin{array}{c}
                  \Diamond_b \\ -1
                 \end{array}\right)
+ \sum_{n=1}^{h}   
\left(\begin{array}{c}
                  \Diamond_a^{n} \\ n-1
                 \end{array}\right)                 
                 \, .
\end{equation}
where $h$ is the height of $\Diamond_a$ below $F$ (takes as a positive number) and $\Diamond_a^{n}$ are all lattice points on $\Diamond_a$ at height $n$. The decomposition of $\Diamond_a$ into the $\Diamond_a^{n}$ is forced on us by demanding that $\langle m, \nu \rangle \geq -1$. Note that $\Diamond_a$ sits below $F$, which leads to a non-positive inner product of $(\Diamond_a,0)$ with $\zeta_a$. In particular, $\langle (\Diamond_a^n,0), \zeta_a \rangle = -n$.

First note that for any non-zero $\zeta_a \zeta_b$ in a small disk around $\zeta_a \zeta_b = 0$ we get a smooth Calabi-Yau threefold $X_{(\Delta,\Delta^\circ)}$. Let us now investigate the geometry of the central fibre, which splits into the two components $\zeta_a=$ and $\zeta_b=0$.

Let us first consider $\zeta_b = 0$. Due to the $SR$-ideal of $\mathbb{P}_\Sigma^\zeta$, this means we can set all coordinates $\nu$ corresponding to lattice points on $\Diamond^\circ_a$ which are not on $\Delta^\circ_F$ to $1$. Furthermore, for any $m$ such that
\begin{equation}
 \langle\, m, (0,0,0,0,1)\,\rangle > -1
\end{equation}
the corresponding monomial in \eqref{eq:familyXdeglimit} vanishes. We hence find
\begin{equation}
 P|_{\zeta_b = 0} = \sum_{m \in \Diamond_b} c_m \prod_{\nu \in \Diamond^\circ_b} z_\nu^{\langle m,\nu \rangle + 1} \zeta_a^{\langle m, (0,0,0,1)\rangle} \, .
\end{equation}
which is precisely the defining equation of $Z_{\Diamond_b,\Diamond^\circ_b}$, where $\zeta_a$ plays the role of $\nu_0$. 

Similarly, for $\zeta_a = 0$, we can set all coordinates $\nu$ corresponding to lattice points on $\Diamond^\circ_b$ which are not on $\Delta^\circ_F$ to $1$ and for any $m$ such that
\begin{equation}
 \langle\, m, (0,0,0,1,1)\,\rangle > -1
\end{equation}
the corresponding monomial in \eqref{eq:familyXdeglimit} vanishes. This means that 
\begin{equation}
\begin{aligned}
P|_{\zeta_a = 0} & =  \sum_{n = 0}^h \sum_{m \in \Diamond_a^n} \prod_{\nu \in \Diamond^\circ_a} z_\nu^{\langle m, \nu \rangle + 1} \zeta_b^n \\
 & = \sum_{m \in \Diamond_a} \prod_{\nu \in \Diamond^\circ_a} z_\nu^{\langle m, \nu \rangle + 1} \zeta_b^{\langle m, \nu_0\rangle } \, ,
\end{aligned}
\end{equation}
so we also recover $Z_{\Diamond_a,\Diamond^\circ_a}$, with $\zeta_b$ playing the role of $\nu_0$.  

Let us perform the degeneration limit for a simple example. Consider the reflexive polytope constructed from two copies of the top used in Section \ref{sect:example_Quartic}, 
\begin{equation}
\Diamond^\circ_a = \left(\begin{array}{rrrrr}
-1 & 0 & 0 & 0 & 1 \\
-1 & 0 & 0 & 1 & 0 \\
-1 & 0 & 1 & 0 & 0 \\
0 & 1 & 0 & 0 & 0
\end{array}\right) \, , \hspace{1cm}
\Diamond^\circ_b = \left(\begin{array}{rrrrr}
-1 & 0 & 0 & 0 & 1 \\
-1 & 0 & 0 & 1 & 0 \\
-1 & 0 & 1 & 0 & 0 \\
0 & -1 & 0 & 0 & 0
\end{array}\right)
\end{equation}
The corresponding Calabi-Yau threefold is simply given by a hypersurface of bidegree $(4,2)$ in $\P^3 \times \P^1$. Denoting the homogeneous coordinates of the $\P^1$ by $[z_1:z_2]$ and the homogeneous coordinates of the $\P^3$ by $[x_1:x_2:x_3:x_4]$, the defining equation of the family $\mathcal{X}_\zeta$ is of the form
\begin{equation}
z_1^2\left(\zeta_a P(x) + \zeta_a^2 \zeta_b Q(x) + \zeta_a^3 \zeta_b^2 R(x) \right) + z_1 z_2 \left(S(x) + \zeta_a \zeta_b T(x) + \zeta_a^2 \zeta_b^2 U(x) \right) + z_2^2 \zeta_b V(x) \, ,
\end{equation}
where $P(x),Q(x),\cdots$ are homogeneous polynomials in the $x_i$ of degree $4$. The Stanley-Reisner ideal of $\P_\Sigma^\zeta$ is simply given by 
$(\zeta_a,z_2)$ and $(\zeta_b,z_1)$. Hence we find that $\zeta_a = 0$ is simply given by
\begin{equation}
z_1 S(x) + \zeta_b V(x) = 0 
\end{equation}
in $\P^3 \times \P^1$ where now the $\P^1$ has homogeneous coordinates $[z_1:\zeta_b]$, so that we recover the canonical form for $Z_{(\Diamond_a,
\Diamond^\circ_a)}$. Similarly, $Z_{(\Diamond_b,\Diamond^\circ_b)}$ is given by $\zeta_b=0$ which gives
\begin{equation}
z_2 S(x) + \zeta_a P(x) =0 
\end{equation}
in $\P^3 \times \P^1$ where now the $\P^1$ has homogeneous coordinates $[z_2:\zeta_a]$. Note that the constant fibre along the bulk region of the base of the $K3$ fibred Calabi-Yau threefold in the limit $\zeta_a\zeta_b \rightarrow 0$, which is $\zeta_a = \zeta_b=0$, is given by the quartic $K3$ surface
\begin{equation}
 S(x) = 0
\end{equation}
in $\P^3$.

A generic hypersurface in $\P^3\times \P^1$ of bidegree $(4,n)$ has 
\begin{equation}
 n \cdot 4 \cdot 3^3 
\end{equation}
singular fibres over which the fibre has an $A_1$ singularity \cite{GKZ}. A Calabi-Yau hypersurface in $\P^3\times \P^1$ hence has $216$ singular fibres and its Euler characteristic is correspondingly given by
\begin{equation}
 (2-216)\cdot 24 + 216 \cdot 23 = -168 \, .
\end{equation} 
The building blocks $Z_{(\Diamond_a,\Diamond^\circ_a)}$ and $Z_{(\Diamond_b,\Diamond^\circ_b)}$ hence each have $108$ singular fibres and their Euler characteristics satisfies (compare with the Hodge numbers computed in Section \ref{sect:example_Quartic}):
\begin{equation} 
 (2-108)\cdot 24 + 108 \cdot 23 = -60 \, .
\end{equation}
In the degeneration limit, the $216$ singular fibres are distributed into $2\cdot 108$ fibres which move towards the ends of the elongated $\P^1$. Note that these Euler characteristics satisfy
\begin{equation}
 \chi(X_{(\Delta,\Delta^\circ)}) = \chi(Z_{(\Diamond_a,\Diamond^\circ_a)}) -24 + \chi(Z_{(\Diamond_b,\Diamond^\circ_b)}) - 24 \, ,
\end{equation}
which fits with the fact that we can cut $X_{(\Delta,\Delta^\circ)}$ into  $\left( Z_{(\Diamond_a,\Diamond^\circ_a)}\setminus S_{0}\right) \amalg  \left(Z_{(\Diamond_b,\Diamond^\circ_b)}\setminus S_{0} \right) \amalg (S_0 \times S^1)$. We expect such relations to hold in complete generality, but are not going to prove them here. 

\section{Mirror Symmetry for K3 Surfaces}\label{sect:mirrorfork3}

In this Section, we review some aspects of mirror map for algebraic $K3$ surfaces \cite{Aspinwall:1994rg,Nahm:2001kh}. The Teichm\"uller moduli space of Ricci-flat metrics on $K3$ surfaces is given by the Grassmanian
\begin{equation}
 \mathcal{T}^{3,19} =  \frac{O(3,19)}{O(3)\times O(19)} \, ,
\end{equation}
times $\R_+$ representing the volume. The threeplane $\Sigma_3$ appearing in this Grassmanian can be thought of as being spanned by the K\"ahler form $\omega$ and the real and imaginary parts of the holomorphic two-form $\Omega$. 

The Teichm\"uller moduli space of $N=(4,4)$ $K3$ $\sigma$-models is a also a Grassmanian, this time of four-planes $\Sigma_4$ in $\Gamma^{4,20}\otimes \R$
\cite{Aspinwall:1994rg,Nahm:2001kh}
\begin{equation}
\mathcal{T}^{4,20} =  \frac{O(4,20)}{O(4)\times O(20)} \, .
\end{equation}
This space is isomorphic to
\begin{equation}\label{eq:isomsigma_geom}
\mathcal{T}^{4,20} \cong  \mathcal{T}^{3,19} \times \R_+ \times \R^{3,19} \,.
\end{equation}
Here, the first factor are the geometric moduli of the $K3$ surface, the second factor is the volume of the $K3$ form, and the third factor is the $B$-field, which takes values in $H^{2}(K3,\R)$. As we are interested in algebraic $K3$ surfaces, we will fix $\Omega$ in $\Sigma_3$ making a choice of complex structure. 

The explicit form of the above isomorphism \eqref{eq:isomsigma_geom} depends on a choice of embedding of $U$ in the unique even unimodular lattice $\Gamma^{4,20}$ which is called a geometric interpretation of the $\sigma$-model. Let us denote the generators of $U_N=H^{0}(K3,\Z)\oplus H^{4}(K3,\Z) $ by $v_0$ and $v$. With a choice of complex structure, the explicit isomorphism is then given by 
\begin{equation}\label{eq:threetofour}
\begin{aligned}
\hat{\Omega} & = \Omega - (\Omega \cdot B) v \\
\hat{\omega} & = \omega - (\omega \cdot B) v \\
\hat{B} & =  B + v_0 +  \tfrac12(\omega\cdot \omega - B\cdot B) v
\end{aligned}
\end{equation}
as the vectors in $\Gamma^{4,20}$ spanning $\Sigma_4$. For algebraic $K3$ surfaces, it is natural to furthermore require that $\Omega \cdot B = 0$, so $\hat{\Omega} = \Omega$ sits purely in $\Gamma^{3,19}$. In this case we may use that $\Gamma^{3,19} = U_T \oplus \Gamma^{2,18}$ and exploit the fact that for $z \in \C$ $\Omega$ and $z \Omega$ give equivalent complex structures to choose a parametrization 
\begin{equation}\label{eq:twotothree}
\begin{aligned}
Re(\Omega) & = Re(\Omega)_2 - (Re(\Omega)_2\cdot Im(\Omega)_2) w \\
Im(\Omega) & = Im(\Omega)_2 + w_0 +  \tfrac12((Re(\Omega)_2)^2-(Im(\Omega)_2)^2) w \, .
\end{aligned}
\end{equation}
Here $U_T$ is spanned by $w_0$ and $w$ and $ Re(\Omega)_2$ and $Im(\Omega)_2$ denote the projections to $\Gamma^{2,18}\otimes\R$. 

Mirror symmetry for $K3$ surfaces can be formulated in terms of an automorphism of the lattice $\Gamma^{4,20}$ which identifies the two-plane spanned by $\hat{B}$ and $\hat{\omega}$ with that spanned by the real and imaginary parts of $\hat{\Omega}$. This is equivalent to choosing different geometric interpretations. For a given geometric interpretation $U_N \hookrightarrow \Gamma^{4,20}$, we must have $\Sigma \perp v$ and, for algebraic $K3$ surfaces,
$B \perp \Omega$. 
If we exchange $U_N$ with $U_T$ we hence arrive at a new geometric interpretation with
\begin{equation}\label{eq:k3mirrorswap}
\begin{aligned}
& Re(\Omega)_2^\circ  = \omega  &Im(\Omega)_2^\circ & = B \\
&\omega^\circ  = Re(\Omega)_2 &B^\circ & = Im(\Omega)_2.\\
\end{aligned}
\end{equation}
This connects two different geometric interpretation which correspond to the same point in the moduli space of the $\sigma$-model. Note that for algebraic families, we wish to furthermore exchange the complex structure moduli with the K\"ahler moduli, so that we are led to consider a pair of lattices $\tilde{T}$ and $N$ with primitive embeddings
\begin{equation}
 \begin{aligned}
\tilde{T}  \hookrightarrow \Gamma^{2,18} \, ,\hspace{1cm} N \hookrightarrow \Gamma^{2,18} 
\end{aligned}
\end{equation}
and $N = \tilde{T}^\perp$, which are exchanged under mirror symmetry. Here $N$ is the polarizing lattice and $T = U_T \oplus \tilde{T}$ is the transcendental lattice of the (generic member of the family of) lattice polarized $K3$ surfaces under consideration.

For $K3$ surfaces which are toric hypersurfaces $X_{(\Delta,\Delta^\circ)}$, such a pair of lattices is found as $\mbox{Pic}_{tor}(X_{\Delta,\Delta^\circ})$ and $n_{(\Delta_F,\Delta^\circ_F)}$ \cite{Rohsiepe:2004st}, or, more generally $N(Z_{(\Diamond,\Diamond^\circ)})$ and $N(Z_{(\Diamond^\circ,\Diamond)})$ for a dual pair of projecting tops (see Appendix \ref{sect:mirrorbuildblocksdetails}).

\bibliography{G2MIR.bib}

\providecommand{\href}[2]{#2}\begingroup\raggedright\begin{thebibliography}{10}

\bibitem{Dixon:1987bg}
L.~J. Dixon, ``{Some world sheet properties of superstring compactifications,
  on orbifolds and otherwise},'' in {\em {Proceedings, Summer Workshop in
  High-energy Physics and Cosmology: Superstrings, Unified Theories and
  Cosmology: Trieste, Italy, June 29-August 7, 1987}}.
\newblock
1987.
\newblock

\bibitem{Lerche:1989uy}
W.~Lerche, C.~Vafa, and N.~P. Warner, ``{Chiral Rings in N=2 Superconformal
  Theories},''
\href{http://dx.doi.org/10.1016/0550-3213(89)90474-4}{{\em Nucl. Phys.}
  {\bfseries B324} (1989) 427--474}.

\bibitem{Candelas:1989hd}
P.~Candelas, M.~Lynker, and R.~Schimmrigk, ``{Calabi-Yau Manifolds in Weighted
  P(4)},''
\href{http://dx.doi.org/10.1016/0550-3213(90)90185-G}{{\em Nucl. Phys.}
  {\bfseries B341} (1990) 383--402}.

\bibitem{Greene:1990ud}
B.~R. Greene and M.~R. Plesser, ``{Duality in {Calabi-Yau} Moduli Space},''
\href{http://dx.doi.org/10.1016/0550-3213(90)90622-K}{{\em Nucl. Phys.}
  {\bfseries B338} (1990) 15--37}.

\bibitem{Strominger:1995cz}
A.~Strominger, ``{Massless black holes and conifolds in string theory},''
  \href{http://dx.doi.org/10.1016/0550-3213(95)00287-3}{{\em Nucl. Phys.}
  {\bfseries B451} (1995) 96--108},
\href{http://arxiv.org/abs/hep-th/9504090}{{\ttfamily arXiv:hep-th/9504090
  [hep-th]}}.

\bibitem{Aspinwall:1995td}
P.~S. Aspinwall and D.~R. Morrison, ``{U duality and integral structures},''
  \href{http://dx.doi.org/10.1016/0370-2693(95)00745-7}{{\em Phys. Lett.}
  {\bfseries B355} (1995) 141--149},
\href{http://arxiv.org/abs/hep-th/9505025}{{\ttfamily arXiv:hep-th/9505025
  [hep-th]}}.

\bibitem{Morrison:1995yi}
D.~R. Morrison, ``{Mirror symmetry and the type II string},''
  \href{http://dx.doi.org/10.1016/0920-5632(96)00016-3}{{\em Nucl. Phys. Proc.
  Suppl.} {\bfseries 46} (1996) 146--155},
\href{http://arxiv.org/abs/hep-th/9512016}{{\ttfamily arXiv:hep-th/9512016
  [hep-th]}}.

\bibitem{Strominger:1996it}
A.~Strominger, S.-T. Yau, and E.~Zaslow, ``{Mirror symmetry is T duality},''
  \href{http://dx.doi.org/10.1016/0550-3213(96)00434-8}{{\em Nucl. Phys.}
  {\bfseries B479} (1996) 243--259},
\href{http://arxiv.org/abs/hep-th/9606040}{{\ttfamily arXiv:hep-th/9606040
  [hep-th]}}.

\bibitem{Becker:1996ay}
K.~Becker, M.~Becker, D.~R. Morrison, H.~Ooguri, Y.~Oz, and Z.~Yin,
  ``{Supersymmetric cycles in exceptional holonomy manifolds and Calabi-Yau 4
  folds},'' \href{http://dx.doi.org/10.1016/S0550-3213(96)00491-9}{{\em Nucl.
  Phys.} {\bfseries B480} (1996) 225--238},
\href{http://arxiv.org/abs/hep-th/9608116}{{\ttfamily arXiv:hep-th/9608116
  [hep-th]}}.

\bibitem{Shatashvili:1994zw}
S.~L. Shatashvili and C.~Vafa, ``{Superstrings and manifold of exceptional
  holonomy},'' \href{http://dx.doi.org/10.1007/BF01671569}{{\em Selecta Math.}
  {\bfseries 1} (1995) 347},
\href{http://arxiv.org/abs/hep-th/9407025}{{\ttfamily arXiv:hep-th/9407025
  [hep-th]}}.

\bibitem{Papadopoulos:1995da}
G.~Papadopoulos and P.~K. Townsend, ``{Compactification of D = 11 supergravity
  on spaces of exceptional holonomy},''
  \href{http://dx.doi.org/10.1016/0370-2693(95)00929-F}{{\em Phys. Lett.}
  {\bfseries B357} (1995) 300--306},
\href{http://arxiv.org/abs/hep-th/9506150}{{\ttfamily arXiv:hep-th/9506150
  [hep-th]}}.

\bibitem{Acharya:1996fx}
B.~S. Acharya, ``{Dirichlet Joyce manifolds, discrete torsion and duality},''
  \href{http://dx.doi.org/10.1016/S0550-3213(97)00163-6}{{\em Nucl. Phys.}
  {\bfseries B492} (1997) 591--606},
\href{http://arxiv.org/abs/hep-th/9611036}{{\ttfamily arXiv:hep-th/9611036
  [hep-th]}}.

\bibitem{Acharya:1997rh}
B.~S. Acharya, ``{On mirror symmetry for manifolds of exceptional holonomy},''
  \href{http://dx.doi.org/10.1016/S0550-3213(98)00140-0}{{\em Nucl. Phys.}
  {\bfseries B524} (1998) 269--282},
\href{http://arxiv.org/abs/hep-th/9707186}{{\ttfamily arXiv:hep-th/9707186
  [hep-th]}}.

\bibitem{Figueroa-OFarrill:1996tnk}
J.~M. Figueroa-O'Farrill, ``{A Note on the extended superconformal algebras
  associated with manifolds of exceptional holonomy},''
  \href{http://dx.doi.org/10.1016/S0370-2693(96)01506-7}{{\em Phys. Lett.}
  {\bfseries B392} (1997) 77--84},
\href{http://arxiv.org/abs/hep-th/9609113}{{\ttfamily arXiv:hep-th/9609113
  [hep-th]}}.

\bibitem{Howe:1991im}
P.~S. Howe and G.~Papadopoulos, ``{W symmetries of a class of d = 2 N=1
  supersymmetric sigma models},''
\href{http://dx.doi.org/10.1016/0370-2693(91)90946-N}{{\em Phys. Lett.}
  {\bfseries B267} (1991) 362--365}.

\bibitem{Howe:1991vs}
P.~S. Howe and G.~Papadopoulos, ``{A Note on holonomy groups and sigma
  models},''
\href{http://dx.doi.org/10.1016/0370-2693(91)90591-D}{{\em Phys. Lett.}
  {\bfseries B263} (1991) 230--232}.

\bibitem{Howe:1991ic}
P.~S. Howe and G.~Papadopoulos, ``{Holonomy groups and W symmetries},''
  \href{http://dx.doi.org/10.1007/BF02097022}{{\em Commun. Math. Phys.}
  {\bfseries 151} (1993) 467--480},
\href{http://arxiv.org/abs/hep-th/9202036}{{\ttfamily arXiv:hep-th/9202036
  [hep-th]}}.

\bibitem{Howe:1994tv}
P.~S. Howe, G.~Papadopoulos, and P.~C. West, ``{Free fermions and extended
  conformal algebras},''
  \href{http://dx.doi.org/10.1016/0370-2693(94)90635-1}{{\em Phys. Lett.}
  {\bfseries B339} (1994) 219--222},
\href{http://arxiv.org/abs/hep-th/9407183}{{\ttfamily arXiv:hep-th/9407183
  [hep-th]}}.

\bibitem{Odake:1988bh}
S.~Odake, ``{Extension of $N=2$ Superconformal Algebra and Calabi-yau
  Compactification},''
\href{http://dx.doi.org/10.1142/S021773238900068X}{{\em Mod. Phys. Lett.}
  {\bfseries A4} (1989) 557}.

\bibitem{Roiban:2002iv}
R.~Roiban, C.~Romelsberger, and J.~Walcher, ``{Discrete torsion in singular
  G(2) manifolds and real LG},'' {\em Adv. Theor. Math. Phys.} {\bfseries 6}
  (2003) 207--278,
\href{http://arxiv.org/abs/hep-th/0203272}{{\ttfamily arXiv:hep-th/0203272
  [hep-th]}}.

\bibitem{Gaberdiel:2004vx}
M.~R. Gaberdiel and P.~Kaste, ``{Generalized discrete torsion and mirror
  symmetry for g(2) manifolds},''
  \href{http://dx.doi.org/10.1088/1126-6708/2004/08/001}{{\em JHEP} {\bfseries
  08} (2004) 001},
\href{http://arxiv.org/abs/hep-th/0401125}{{\ttfamily arXiv:hep-th/0401125
  [hep-th]}}.

\bibitem{Zamolodchikov:1986gt}
A.~B. Zamolodchikov, ``{Irreversibility of the Flux of the Renormalization
  Group in a 2D Field Theory},'' {\em JETP Lett.} {\bfseries 43} (1986)
  730--732.
[Pisma Zh. Eksp. Teor. Fiz.43,565(1986)].

\bibitem{Candelas:1989qn}
P.~Candelas, T.~Hubsch, and R.~Schimmrigk, ``{Relation Between the
  Weil-petersson and Zamolodchikov Metrics},''
\href{http://dx.doi.org/10.1016/0550-3213(90)90072-L}{{\em Nucl. Phys.}
  {\bfseries B329} (1990) 583--590}.

\bibitem{MR2024648}
A.~Kovalev, ``Twisted connected sums and special {R}iemannian holonomy,''
  \href{http://dx.doi.org/10.1515/crll.2003.097}{{\em J. Reine Angew. Math.}
  {\bfseries 565} (2003) 125--160}.
  \url{http://dx.doi.org/10.1515/crll.2003.097}.

\bibitem{Corti:2012kd}
A.~Corti, M.~Haskins, J.~Nordström, and T.~Pacini,
  ``{$\mathrm{G}_{2}$-manifolds and associative submanifolds via semi-Fano
  $3$-folds},'' \href{http://dx.doi.org/10.1215/00127094-3120743}{{\em Duke
  Math. J.} {\bfseries 164} no.~10, (2015) 1971--2092},
\href{http://arxiv.org/abs/1207.4470}{{\ttfamily arXiv:1207.4470 [math.DG]}}.

\bibitem{Halverson:2014tya}
J.~Halverson and D.~R. Morrison, ``{The landscape of M-theory compactifications
  on seven-manifolds with G$_{2}$ holonomy},''
  \href{http://dx.doi.org/10.1007/JHEP04(2015)047}{{\em JHEP} {\bfseries 04}
  (2015) 047},
\href{http://arxiv.org/abs/1412.4123}{{\ttfamily arXiv:1412.4123 [hep-th]}}.

\bibitem{Halverson:2015vta}
J.~Halverson and D.~R. Morrison, ``{On gauge enhancement and singular limits in
  G$_{2}$ compactifications of M-theory},''
  \href{http://dx.doi.org/10.1007/JHEP04(2016)100}{{\em JHEP} {\bfseries 04}
  (2016) 100},
\href{http://arxiv.org/abs/1507.05965}{{\ttfamily arXiv:1507.05965 [hep-th]}}.

\bibitem{Batyrev:1994hm}
V.~V. Batyrev, ``{Dual polyhedra and mirror symmetry for Calabi-Yau
  hypersurfaces in toric varieties},'' {\em J. Alg. Geom.} {\bfseries 3} (1994)
  493--545,
\href{http://arxiv.org/abs/alg-geom/9310003}{{\ttfamily arXiv:alg-geom/9310003
  [alg-geom]}}.

\bibitem{Batyrev:1994pg}
V.~V. Batyrev and L.~A. Borisov, ``{On Calabi-Yau complete intersections in
  toric varieties},''
\href{http://arxiv.org/abs/alg-geom/9412017}{{\ttfamily arXiv:alg-geom/9412017
  [alg-geom]}}.

\bibitem{Braun:2016igl}
A.~P. Braun, ``{Tops as Building Blocks for G2 Manifolds},''
\href{http://arxiv.org/abs/1602.03521}{{\ttfamily arXiv:1602.03521 [hep-th]}}.

\bibitem{Aspinwall:1990ck}
P.~S. Aspinwall and C.~A. Lutken, ``{Quantum algebraic geometry of superstring
  compactifications},''
\href{http://dx.doi.org/10.1016/0550-3213(91)90123-F}{{\em Nucl. Phys.}
  {\bfseries B355} (1991) 482--510}.

\bibitem{Gross:1997hn}
M.~Gross, ``{Special Lagrangian fibrations. I: Topology},'' in {\em {Integrable
  systems and algebraic geometry. Proceedings, Taniguchi Symposium, Kobe,
  Japan, June 30-July 4, 1997, Kyoto, Japan, July 7-July 11, 1997}},
  pp.~156--193.
\newblock
1997.
\newblock

\bibitem{Batyrev:2005jc}
V.~Batyrev and M.~Kreuzer, ``{Integral cohomology and mirror symmetry for
  Calabi-Yau 3-folds},''
\href{http://arxiv.org/abs/math/0505432}{{\ttfamily arXiv:math/0505432
  [math-ag]}}.

\bibitem{Witten:1993yc}
E.~Witten, ``{Phases of N=2 theories in two-dimensions},''
  \href{http://dx.doi.org/10.1016/0550-3213(93)90033-L}{{\em Nucl. Phys.}
  {\bfseries B403} (1993) 159--222},
\href{http://arxiv.org/abs/hep-th/9301042}{{\ttfamily arXiv:hep-th/9301042
  [hep-th]}}.

\bibitem{Hori:2000kt}
K.~Hori and C.~Vafa, ``{Mirror symmetry},''
\href{http://arxiv.org/abs/hep-th/0002222}{{\ttfamily arXiv:hep-th/0002222
  [hep-th]}}.

\bibitem{Aganagic:2002mp}
M.~Aganagic and C.~Vafa, ``{Perturbative derivation of mirror symmetry},''
\href{http://arxiv.org/abs/hep-th/0209138}{{\ttfamily arXiv:hep-th/0209138
  [hep-th]}}.

\bibitem{deBoer:2005pt}
J.~de~Boer, A.~Naqvi, and A.~Shomer, ``{The Topological G(2) string},''
  \href{http://dx.doi.org/10.4310/ATMP.2008.v12.n2.a2}{{\em Adv. Theor. Math.
  Phys.} {\bfseries 12} no.~2, (2008) 243--318},
\href{http://arxiv.org/abs/hep-th/0506211}{{\ttfamily arXiv:hep-th/0506211
  [hep-th]}}.

\bibitem{Acharya:2000gb}
B.~S. Acharya, ``{On Realizing N=1 superYang-Mills in M theory},''
\href{http://arxiv.org/abs/hep-th/0011089}{{\ttfamily arXiv:hep-th/0011089
  [hep-th]}}.

\bibitem{Atiyah:2001qf}
M.~Atiyah and E.~Witten, ``{M theory dynamics on a manifold of G(2)
  holonomy},'' {\em Adv. Theor. Math. Phys.} {\bfseries 6} (2003) 1--106,
\href{http://arxiv.org/abs/hep-th/0107177}{{\ttfamily arXiv:hep-th/0107177
  [hep-th]}}.

\bibitem{Witten:2001uq}
E.~Witten, ``{Anomaly cancellation on G(2) manifolds},''
\href{http://arxiv.org/abs/hep-th/0108165}{{\ttfamily arXiv:hep-th/0108165
  [hep-th]}}.

\bibitem{Acharya:2001gy}
B.~S. Acharya and E.~Witten, ``{Chiral fermions from manifolds of G(2)
  holonomy},''
\href{http://arxiv.org/abs/hep-th/0109152}{{\ttfamily arXiv:hep-th/0109152
  [hep-th]}}.

\bibitem{Gukov:2002es}
S.~Gukov and D.~Tong, ``{D-brane probes of special holonomy manifolds, and
  dynamics of N = 1 three-dimensional gauge theories},''
  \href{http://dx.doi.org/10.1088/1126-6708/2002/04/050}{{\em JHEP} {\bfseries
  04} (2002) 050},
\href{http://arxiv.org/abs/hep-th/0202126}{{\ttfamily arXiv:hep-th/0202126
  [hep-th]}}.

\bibitem{Acharya:2004qe}
B.~S. Acharya and S.~Gukov, ``{M theory and singularities of exceptional
  holonomy manifolds},''
  \href{http://dx.doi.org/10.1016/j.physrep.2003.10.017}{{\em Phys. Rept.}
  {\bfseries 392} (2004) 121--189},
\href{http://arxiv.org/abs/hep-th/0409191}{{\ttfamily arXiv:hep-th/0409191
  [hep-th]}}.

\bibitem{Aganagic:2001ug}
M.~Aganagic and C.~Vafa, ``{G(2) manifolds, mirror symmetry and geometric
  engineering},''
\href{http://arxiv.org/abs/hep-th/0110171}{{\ttfamily arXiv:hep-th/0110171
  [hep-th]}}.

\bibitem{Gukov:2002jv}
S.~Gukov, S.-T. Yau, and E.~Zaslow, ``{Duality and fibrations on G(2)
  manifolds},''
\href{http://arxiv.org/abs/hep-th/0203217}{{\ttfamily arXiv:hep-th/0203217
  [hep-th]}}.

\bibitem{MR3109862}
A.~Corti, M.~Haskins, J.~Nordstr{\"o}m, and T.~Pacini, ``Asymptotically
  cylindrical {C}alabi-{Y}au 3-folds from weak {F}ano 3-folds,''
  \href{http://dx.doi.org/10.2140/gt.2013.17.1955}{{\em Geom. Topol.}
  {\bfseries 17} no.~4, (2013) 1955--2059}.
  \url{http://dx.doi.org/10.2140/gt.2013.17.1955}.

\bibitem{Harvey:1982xk}
R.~Harvey and H.~B. Lawson, Jr., ``{Calibrated geometries},''
\href{http://dx.doi.org/10.1007/BF02392726}{{\em Acta Math.} {\bfseries 148}
  (1982) 47}.

\bibitem{Mclean96}
R.~C. Mclean, ``Deformations of calibrated submanifolds,'' {\em Commun. Analy.
  Geom} {\bfseries 6} (1996) 705--747.

\bibitem{Aspinwall:1996mn}
P.~S. Aspinwall, ``{K3 surfaces and string duality},'' in {\em {Fields, strings
  and duality. Proceedings, Summer School, Theoretical Advanced Study Institute
  in Elementary Particle Physics, TASI'96, Boulder, USA, June 2-28, 1996}},
  pp.~421--540.
\newblock 1996.
\newblock
\href{http://arxiv.org/abs/hep-th/9611137}{{\ttfamily arXiv:hep-th/9611137
  [hep-th]}}.
\newblock

\bibitem{Morrison:2010vf}
D.~R. Morrison, ``{On the structure of supersymmetric T**3 fibrations},''
\href{http://arxiv.org/abs/1002.4921}{{\ttfamily arXiv:1002.4921 [math.AG]}}.

\bibitem{Gross:2012rw}
M.~Gross, ``{Mirror symmetry and the Strominger-Yau-Zaslow conjecture},''
  \href{http://dx.doi.org/10.4310/CDM.2012.v2012.n1.a3}{{\em Current
  Developments in Mathematics} {\bfseries 1} (2012) 133--191},
  \href{http://arxiv.org/abs/1212.4220}{{\ttfamily arXiv:1212.4220 [math.AG]}}.

\bibitem{Earp:2013jea}
H.~Sa~Earp and T.~Walpuski, ``{${\rm G}_2$-instantons on twisted connected
  sums},''
\href{http://arxiv.org/abs/1310.7933}{{\ttfamily arXiv:1310.7933 [math.DG]}}.

\bibitem{MR873655}
V.~I. Danilov and A.~G. Khovanskii, ``Newton polyhedra and an algorithm for
  calculating {H}odge-{D}eligne numbers,'' {\em Izv. Akad. Nauk SSSR Ser. Mat.}
  {\bfseries 50} no.~5, (1986) 925--945.

\bibitem{Klemm:1995tj}
A.~Klemm, W.~Lerche, and P.~Mayr, ``{K3 Fibrations and heterotic type II string
  duality},'' \href{http://dx.doi.org/10.1016/0370-2693(95)00937-G}{{\em Phys.
  Lett.} {\bfseries B357} (1995) 313--322},
\href{http://arxiv.org/abs/hep-th/9506112}{{\ttfamily arXiv:hep-th/9506112
  [hep-th]}}.

\bibitem{Candelas:1996su}
P.~Candelas and A.~Font, ``{Duality between the webs of heterotic and type II
  vacua},'' \href{http://dx.doi.org/10.1016/S0550-3213(96)00410-5}{{\em
  Nucl.Phys.} {\bfseries B511} (1998) 295--325},
\href{http://arxiv.org/abs/hep-th/9603170}{{\ttfamily arXiv:hep-th/9603170
  [hep-th]}}.

\bibitem{Avram:1996pj}
A.~C. Avram, M.~Kreuzer, M.~Mandelberg, and H.~Skarke, ``{Searching for K3
  fibrations},'' \href{http://dx.doi.org/10.1016/S0550-3213(97)00214-9}{{\em
  Nucl. Phys.} {\bfseries B494} (1997) 567--589},
\href{http://arxiv.org/abs/hep-th/9610154}{{\ttfamily arXiv:hep-th/9610154
  [hep-th]}}.

\bibitem{Kreuzer:2000qv}
M.~Kreuzer and H.~Skarke, ``{Reflexive polyhedra, weights and toric Calabi-Yau
  fibrations},'' \href{http://dx.doi.org/10.1142/S0129055X0200120X}{{\em Rev.
  Math. Phys.} {\bfseries 14} (2002) 343--374},
\href{http://arxiv.org/abs/math/0001106}{{\ttfamily arXiv:math/0001106
  [math-ag]}}.

\bibitem{Candelas:2012uu}
P.~Candelas, A.~Constantin, and H.~Skarke, ``{An Abundance of K3 Fibrations
  from Polyhedra with Interchangeable Parts},''
  \href{http://dx.doi.org/10.1007/s00220-013-1802-2}{{\em Commun. Math. Phys.}
  {\bfseries 324} (2013) 937--959},
\href{http://arxiv.org/abs/1207.4792}{{\ttfamily arXiv:1207.4792 [hep-th]}}.

\bibitem{Aspinwall:1994rg}
P.~S. Aspinwall and D.~R. Morrison, ``{String theory on K3 surfaces},''
\href{http://arxiv.org/abs/hep-th/9404151}{{\ttfamily arXiv:hep-th/9404151
  [hep-th]}}.

\bibitem{MR1420220}
I.~V. Dolgachev, ``Mirror symmetry for lattice polarized {$K3$} surfaces,''
  \href{http://dx.doi.org/10.1007/BF02362332}{{\em J. Math. Sci.} {\bfseries
  81} no.~3, (1996) 2599--2630},
  \href{http://arxiv.org/abs/alg-geom/9502005}{{\ttfamily
  arXiv:alg-geom/9502005 [alg-geom]}}.

\bibitem{Rohsiepe:2004st}
F.~Rohsiepe, ``{Lattice polarized toric K3 surfaces},''
\href{http://arxiv.org/abs/hep-th/0409290}{{\ttfamily arXiv:hep-th/0409290
  [hep-th]}}.

\bibitem{Kreuzer:1998vb}
M.~Kreuzer and H.~Skarke, ``{Classification of reflexive polyhedra in
  three-dimensions},'' {\em Adv. Theor. Math. Phys.} {\bfseries 2} (1998)
  847--864,
\href{http://arxiv.org/abs/hep-th/9805190}{{\ttfamily arXiv:hep-th/9805190
  [hep-th]}}.

\bibitem{MR525944}
V.~V. Nikulin, ``Integer symmetric bilinear forms and some of their geometric
  applications,'' {\em Izv. Akad. Nauk SSSR Ser. Mat.} {\bfseries 43} no.~1,
  (1979) 111--177, 238.

\bibitem{sage}
W.~Stein {\em et~al.}, {\em {S}age {M}athematics {S}oftware ({V}ersion 6.7)}.
\newblock The Sage Development Team, 2015.
\newblock {\tt http://www.sagemath.org}.

\bibitem{1307.6514}
R.~Davis, C.~Doran, A.~Gewiss, A.~Novoseltsev, D.~Skjorshammer, A.~Syryczuk,
  and U.~Whitcher, ``Short tops and semistable degenerations,''
  \href{http://arxiv.org/abs/arXiv:1307.6514}{{\ttfamily arXiv:1307.6514}}.

\bibitem{Cvetic:2015uwu}
M.~Cvetic, A.~Grassi, D.~Klevers, M.~Poretschkin, and P.~Song, ``{Origin of
  Abelian Gauge Symmetries in Heterotic/F-theory Duality},''
  \href{http://dx.doi.org/10.1007/JHEP04(2016)041}{{\em JHEP} {\bfseries 04}
  (2016) 041},
\href{http://arxiv.org/abs/1511.08208}{{\ttfamily arXiv:1511.08208 [hep-th]}}.

\bibitem{GKZ}
I.~M. Gelfand, M.~M. Kapranov, and A.~V. Zelevinsky, {\em {Discriminants,
  Resultants, and Multidimensional Determinants}}.
\newblock Birkhaeuser Boston, 1994.

\bibitem{Nahm:2001kh}
W.~Nahm and K.~Wendland, ``{Mirror symmetry on Kummer type K3 surfaces},''
  \href{http://dx.doi.org/10.1007/s00220-003-0985-3}{{\em Commun. Math. Phys.}
  {\bfseries 243} (2003) 557--582},
\href{http://arxiv.org/abs/hep-th/0106104}{{\ttfamily arXiv:hep-th/0106104
  [hep-th]}}.

\end{thebibliography}\endgroup

\end{document}